\documentclass[10pt, final]{IEEEtran}
\usepackage{amsthm}
\usepackage{amsmath}
\usepackage[ruled]{algorithm2e}
\usepackage{mathrsfs}
\usepackage{cite}
\usepackage{bm,epsfig,amsthm,url}
\usepackage{indentfirst}
\usepackage{amssymb}
\usepackage{amsfonts}
\usepackage{epstopdf}
\usepackage[caption=false,font=footnotesize]{subfig}
\usepackage{xcolor}
\usepackage{mathtools}

\theoremstyle{definition}

\newtheorem{proposition}{Proposition}

\begin{document}

\title{Reconfigurable Intelligent Surface Empowered Downlink Non-Orthogonal Multiple Access}

\author{\IEEEauthorblockN{Min Fu,~\IEEEmembership{Student Member,~IEEE,} Yong Zhou,~\IEEEmembership{Member,~IEEE,} \\ Yuanming Shi,~\IEEEmembership{Member,~IEEE,} and Khaled B. Letaief,~\IEEEmembership{Fellow,~IEEE}}\\
%               ~and Xuemin Shen,~\IEEEmembership{Fellow,~IEEE}}\\
        \thanks{M. Fu  is with the School of Information Science and Technology, ShanghaiTech University, Shanghai 201210, China, and also with the University of Chinese Academy of Sciences, Beijing 100049, (e-mail: fumin@shanghaitech.edu.cn).}
        \thanks{Y. Zhou and Y. Shi are with the School of Information Science and Technology, ShanghaiTech University, Shanghai 201210, China (e-mail:
        \{zhouyong, shiym\}@shanghaitech.edu.cn).}
       \thanks{K. B. Letaief is with the Department of Electronic and Computer Engineering,
           Hong Kong University of Science and Technology, Hong Kong (e-mail:
           eekhaled@ust.hk). He is also with Peng Cheng Laboratory, Shenzhen, China.}
%       \thanks{X. Shen is with the Department of Electrical and Computer Engineering, University of Waterloo, Waterloo, ON N2L 3G1, Canada (e-mail: sshen@uwaterloo.ca).}
        \thanks{This paper has been presented in part at the \textit{IEEE Globecom Workshops}, Waikoloa, Hawaii, Dec. 2019 \cite{Fu2019Intelligent}. }
             }

\maketitle
\vspace{-6mm}
\begin{abstract}
Power-domain non-orthogonal multiple access (NOMA) has become a promising technology to exploit the new dimension of the power domain to enhance the spectral efficiency of wireless networks.
However, most existing NOMA schemes rely on the strong assumption that users' channel gains are quite different,  which may be invalid in practice.
To unleash the potential of power-domain NOMA, 
we will propose a reconfigurable intelligent surface (RIS)-empowered NOMA network to introduce desirable
channel gain differences among the users by adjusting the phase shifts at RIS. Our goal is to minimize the  total transmit power by jointly optimizing the beamforming vectors at the base station  and the phase-shift matrix at the RIS. 
To address the highly coupled optimization variables, we present an alternating optimization framework to decompose the non-convex bi-quadratically constrained quadratic problem into two rank-one constrained matrices optimization problems via matrix lifting. 
At the same time, to accurately detect the feasibility of the non-convex rank-one constraints and improve performance by avoiding early stopping in the alternating optimization procedure, we equivalently represent the rank-one constraint
as the difference between nuclear norm and spectral norm. A difference-of-convex (DC) algorithm is further developed to solve the resulting DC programs
via successive convex relaxation, followed by establishing the convergence of the proposed DC-based
alternating optimization method. 
We further propose an efficient user ordering scheme with closed-form expressions, considering both the channel conditions and users' target data rates. 
Simulation results will validate the effectiveness of deploying an RIS and the superiority of the proposed DC-based alternating optimization method in reducing the total transmit power. 
\end{abstract}

\section{Introduction}
With the upsurge of diversified wireless services and applications such as Internet of Things (IoT) and mobile Internet, various innovative technologies are expected to keep pace with the exponential growth of the mobile data traffic generated by billions of connected devices in the fifth generation (5G) networks and beyond\cite{Andrews2014What, Letaief2019roadmap}.
Therein and to meet the demand of enormous data traffic, the design of appropriate multiple access techniques has been under intense consideration in both academia and industry \cite{Liu2017Nonorthogonal}.
Power-domain non-orthogonal multiple access (NOMA) is recognized as a key enabling  technology that enables the base station (BS) to simultaneously serve multiple users in the same physical resource block (e.g., time and frequency), thereby significantly improving the spectral efficiency and connection density \cite{ Liu2017Nonorthogonal, Dai2015Nonorthogonal, Zhou2018Coverage}. 
 The main idea of downlink power-domain NOMA is that the BS applies superposition coding at the transmitter with the transmit powers as weight factors while each user performing successive interference cancellation (SIC) at the receiver to remove co-channel interference from the received signal before decoding its own signal  \cite{Islam2017Power, Ding2017application}.

 However, most of the existing studies on NOMA assumed that users' channel conditions are quite different\cite{Ding2016MIMONOMA}.
For example, the authors in \cite{Zhou2018Dynamic} assumed the NOMA user pair  includes a user located close to the BS  and the other far from the BS.
The BS then allocates a higher transmit power to the users with worse channel conditions. 
As demonstrated in \cite{Ding2016UserPairing}, an appropriate difference in terms of the users' channel conditions  is important  to unleash the potential of NOMA. 
In particular, the performance gain of NOMA over orthogonal multiple access (OMA) is limited if there are small channel gain differences among the users.
Unfortunately, the simultaneously served users in NOMA networks may not always have diverse and different channel conditions in practical scenarios because this depends on the propagation environment, which are uncontrollable.
For instance, in the internet of vehicles scenarios  \cite{Lu2014Connected},
 multiple users with diversified quality of service (QoS) requirements need to be   simultaneously served by the BS even if  their channel conditions are similar.

Fortunately, with the theoretical and experimental breakthrough in 
micro electromechanical systems and metamaterials (e.g., metasurface) \cite{Cui2014Coding}, a reconfigurable intelligent surface (RIS), as an emerging cost-effective technology, has recently been proposed as a powerful solution to enhance the spectrum-efficiency  and energy-efficiency of wireless networks \cite{Di2019Smart, Yuan2020Reconfigurable, Huang2019Holographic, Liang2019Large}.  
In particular, the RIS is implemented as an array of low-cost scattering elements, each of them being able to induce an adjustable phase shift  to the incident signal to be reflected, thereby  reconfiguring the reflected signal propagations \cite{Wu2019Towards}. 
Different from amplify-and-forward (AF) relay, backscatter communication, and active intelligent surface based massive multiple-input multiple-output (MIMO), the RIS operates in full-duplex mode and only reflects the received signals as a passive array without the need of any transmit radio-frequency (RF) chains to provide spectrum-efficient  and cost-effective communications \cite{Liang2019Large, Wu2019Towards}.
This, thus, motivates the study RIS-empowered downlink NOMA transmission, where the RIS is capable of inducing desirable channel differences among the users to enhance the performance of NOMA.

Energy-efficient communication is a critical design aspect for future wireless networks \cite{Shi2014Group}.
In this paper, we will focus on the power minimization problem for an RIS-empowered downlink multi-user multiple-input single-output (MISO) NOMA network by jointly optimizing both the transmit beamforming vectors at the BS and the phase-shift matrix at the RIS by taking into account the users' target data rates. 
The unique challenges of the power minimization problem arise from  both the non-convex bi-quadratic QoS constraints (due to the highly coupled transmit beamforming vectors and the phase-shift matrix variables) and the non-convex  modulus constraints (due to the RIS hardware setup).
Besides, user ordering has a significant impact on power consumption and is challenging in RIS-empowered multi-user MISO NOMA networks.
Generally, in NOMA networks without RIS, the users are usually ordered based on their channel conditions with respect to the BS. 
 However, in RIS-empowered NOMA networks, the channel condition between the BS and each user depends not only on the direct and reflect channel responses but also on the phase shifts at the RIS. Due to the diverse data rate requirements as well as the combined channel conditions among users, the design of user ordering becomes much more complicated.
 
\subsection{Contributions}
To address the aforementioned unique challenges, we shall develop an effective optimization framework for the total power minimization problem, followed by proposing a low-complexity user ordering method in the RIS-empowered multi-user MISO NOMA networks.  
The main contributions of this paper are summarized as follows.
\begin{itemize}
\item This paper is one of the early attempts to investigate the transmit power minimization problem of RIS-empowered multi-user MISO NOMA networks, where an RIS is deployed to introduce appropriate channel gain differences among users, thereby unleashing the potential of NOMA without relying on the assumption of the diverse channel conditions of the users. Moreover, we further propose an efficient user ordering 
scheme with a closed-form ordering criterion by taking into account the combined channel conditions and the target data rates.

\item To support efficient algorithms design, we adopt the alternating optimization method to decompose the original bi-quadratically constrained quadratic problem into two subproblems with non-convex quadratic constraints, i.e., a  non-convex quadratically constrained quadratic programming (QCQP) subproblem for optimizing transmit beamforming vectors and a non-convex QCQP feasibility subproblem for optimizing the phase-shift matrix. 

\item We develop a unified difference-of-convex (DC) method to solve the aforementioned subproblems with the capability of accurately detecting the feasibility of non-convex quadratic constraints for the transmit beamforming vectors and  phase-shift matrix design, which can avoid early stopping in the alternating optimization procedure, thereby considerably improving the performance compared with state-of-the-art methods.
The main idea is to reformulate the resultant non-convex QCQP subproblems as multiple rank-one constrained matrices optimization problems via matrix lifting, followed by equivalently representing the rank-one constraint as the difference between the nuclear norm and the spectral norm.

\item We further present an efficient  DC algorithm for the resulting non-convex DC programs via successive convex relaxation.
By representing the objective functions of the resultant DC programs as the difference of two strongly convex functions, we prove that the DC algorithm converges to the stationary solution for DC programs and that the DC-based alternating optimization method (namely, alternating DC method) always converges.
\end{itemize}

Simulation results will demonstrate the effectiveness of deploying an RIS and the superiority of the proposed alternating DC method in reducing the total transmit power.
 Besides, the proposed user ordering criterion is a good option for large-size networks, which provides comparable performance to the exhaustive search.

\subsection{Related Works}
\subsubsection{RIS-empowered OMA Networks}
The research on RIS-empowered wireless networks has recently received considerable attention in vast applications, e.g., coverage extension \cite{ Huang2019Reconfigurable},   energy-efficient beamforming \cite{Wu2019intelligentJ,  Guo2019Weighted, Jiang2019Over}, physical layer security \cite{Yu2019Enabling, Chen2019Intelligent}, and massive connectivity \cite{Xia2019Intelligent}, etc.
For coverage extension, the objective of RIS is to create indirect links with the BS and user, in the scenario where there is no direct link between the BS and users, or the direct link is severely blocked by obstacles. Therein,  \cite{Huang2019Reconfigurable} demonstrated that RIS-based networks  achieved higher energy efficiency than conventional AF relays.
The RIS is deployed to assist the efficient beamforming design that compensates for the signal attenuation from the BS or co-channel interference from neighboring BS by jointly optimizing the active beamformer at BS and the phase-shift matrix at RIS. 
In particular, \cite{Wu2019intelligentJ} showed that an RIS-empowered MISO system  significantly reduced the transmit power consumption compared with the system without using an RIS, and achieved the same rate performance as a conventional massive MIMO system, but with considerably decreased active antennas/RF chains. 
\cite{Guo2019Weighted} also showed that the RIS-empowered network outperformed the system without using RIS in terms of the weighted sum rate. 
Besides,  the author in \cite{Jiang2019Over} applied RIS to reduce the distortion between the decoding signal and the ground-truth signal  for over-the-air computation.
The use of RIS for improving physical layer security aims to cancel out the signal from the BS at the eavesdropper. 
Specifically,
\cite{Yu2019Enabling} validated that deploying large-scale RISs  achieved higher significant performance gains than increasing the number of the antenna at the transmitter in terms of secrecy rate
and energy efficiency for a simple scenario with one legitimate receiver and one eavesdropper. 
\cite{Chen2019Intelligent} further confirmed that  the security enhancement provisioning for RIS-empowered networks to keep the signals secret from multiple eavesdroppers in a broadcast system.
 The author in \cite{Xia2019Intelligent} considered an RIS-empowered IoT network to support massive connectivity under the grant-free random access protocol. 
\subsubsection{RIS-empowered NOMA Networks}
Upon the completion of this work, the application of RIS in NOMA networks was investigated in  some parallel works\cite{Ding2019simple, Yang2019Intelligent, Li2019Joint, Zhu2019Power, Mu2019exploiting, Zuo2020Resource}. 
Specifically, the authors in \cite{Ding2019simple} analyzed the transmission reliability of RIS-aided NOMA transmission. 
For a multi-user NOMA network, the authors in \cite{Yang2019Intelligent} optimized the  transmit beamforming vectors at the BS and the phase shifts at RIS to enhance the user fairness.  
In addition, the authors in \cite{Li2019Joint} considered an RIS-empowered downlink MISO NOMA network to minimize the power consumption, where zero-forcing precoding was employed at the BS to cancel the inter-pair interference. 
In \cite{Zhu2019Power}, the authors studied the power efficiency of  RIS-empowered MISO
NOMA system under additional quasi-degraded channels constraints with two users' case.
In \cite{Mu2019exploiting}, the authors investigated  the sum rate maximization problem in RIS-empowered  NOMA networks.
The authors in \cite{Zuo2020Resource} focused on the maximization of the system throughput over the channel assignment, power allocation, and ideal reflection coefficient in a single-input single-output (SISO) NOMA network.

\subsection{Organization}
The remainder of this paper is organized as follows. Section II describes the system model and problem formulation. 
We present an alternating optimization  to decouple the optimization variables in Section III. 
Section IV provides a unified DC  method to reformulate the non-convex QCQP problems into   DC programs. 
In Section V, we present the DC algorithm to solve the resulting DC programs
and prove the convergence of  the  alternating DC algorithm. 
Section VI proposes a low-complexity user ordering scheme with closed-form expressions. 
Section VII  presents the numerical results and Section VIII concludes this paper.

$\textit{Notations}$:  $\mathbb{ C}^{x\times y}$ denotes the space of $x\times y$ complex-valued matrices. $\mathbb{E}(\cdot)$ denotes the statistical expectation. $(\cdot)^{\sf H}$ and $(\cdot)^{\sf T}$   denote the conjugate transpose and transpose, respectively. For a complex-valued vector $\bm x$, $\|\bm x\|$ denotes its Euclidean norm and $\text{diag}(\bm x)$ denotes a diagonal matrix with each diagonal entry being the corresponding element in $\bm x$. 
For a matrix $\bm X$, $\|\bm X\|_F$,  $\|\bm X\|_*$, and $\|\bm X\|_2$ denote its the Frobenius norm, the nuclear norm and the spectral norm, respectively. $j$ denotes the imaginary unit. Finally, $\mathfrak{R}(\cdot)$ denotes the real part of a complex number. 

\section{System Model and Problem Formulation}
\subsection{System Model}
\begin{figure}[t]
	\centering
	\includegraphics[width=9cm, height=6cm]{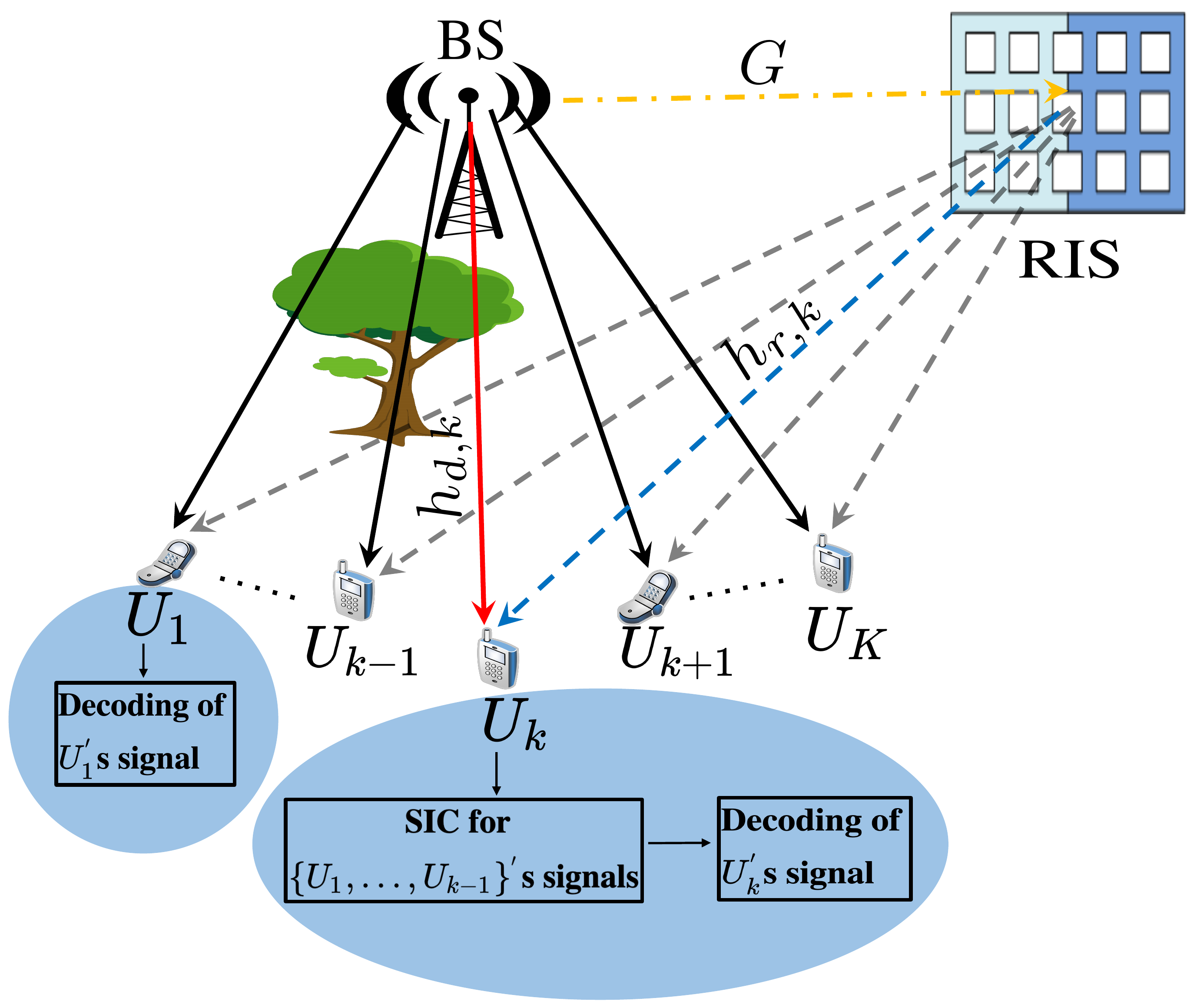}
	\vspace{-4mm}
	\caption{An RIS-empowered downlink MISO NOMA network with $K$ users.    
		%consisting of one BS, $K$ users, and one RIS. 
		User $U_k$ is allocated the $k$-th highest transmit power, and  has to decode and remove the signals intended for users $U_1, \ldots, U_{k-1}$ before decoding its own signal.}\label{systemodel}
	\vspace{-4mm}
\end{figure}
As shown in Fig. \ref{systemodel}, we consider an RIS-empowered multi-user MISO power-domain NOMA system, where an RIS with $N$ passive reflecting elements is deployed to assist the data transmission from  an $M$-antenna BS to $K$ single-antenna users by providing additional channel paths to boost the received signal power and introducing diverse channel conditions between the users and the BS.  
To account for the increasing number of users and the limited spectrum resource, we consider an overloaded scenario, where the number of users is more than the number of antennas at the BS, i.e.,  $K > M$. 
We denote $s_k\in\mathbb{C}$ and $ \bm w_k\in\mathbb{C}^{M} $ as the signal and linear beamforming vector at the BS for user $U_k$, respectively, where $k \in \mathcal{K}$ with $\mathcal{K} = \{1, 2, \ldots, K \}$. 
Without loss of generality, signal $s_k$ is assumed to have zero mean and unit variance, i.e., $\mathbb{E}[s_ks_k^{\sf H}] = 1, \forall \, k \in \mathcal{K}$.
Under universal frequency reuse, the BS superimposes and transmits the signals intended for $K$ users.   After the reflection of the RIS, the signal  received at  user $U_k$ is given by 
\setlength\arraycolsep{2pt}
\begin{eqnarray}\label{receive1}
y_k = (\bm h_{r,k}^{\sf{H}}\bm \Theta\bm G+\bm h_{d,k}^{\sf H})\sum_{j=1}^{K}\bm w_js_j+e_k, \forall \, k \in \mathcal{K},
\end{eqnarray}
where $ \bm h_{d,k}\in\mathbb{C}^M, \bm G\in\mathbb{C}^{N\times M}$, and $\bm h_{r,k}\in\mathbb{C}^{N}$ denote the channel responses from the BS to user $U_k$, from the BS to the RIS, and from the RIS to user $U_k$, respectively. Likewise, $ e_k\sim\mathcal{CN}(0, \sigma^2) $ is the additive white Gaussian noise (AWGN) with $\sigma^2$ being the noise power. Note that quasi-static flat fading model is considered for all channels. 
In addition,  $ \bm\Theta=\text{diag}(\beta e^{j\theta_1},\ldots, \beta e^{j\theta_N}) \in \mathbb{ C}^{N \times N}$ denotes  the diagonal phase-shift matrix of the RIS, where $  \theta_n\in[0,2\pi), \forall n $ and $\beta \in [0,1]$ denote the phase shift of element $n$ and the amplitude reflection coefficient on the incident signal, respectively. 
As each element on the RIS is designed to boost the received signals, we assume $\beta=1$ without loss of generality, similar to \cite{Wu2019intelligentJ,  Huang2019Reconfigurable, Guo2019Weighted, Yu2019Enabling, Chen2019Intelligent, Jiang2019Over}, \cite{Ding2019simple, Yang2019Intelligent, Li2019Joint, Zhu2019Power, Mu2019exploiting}. 
Due to the severe path loss, the power of the signals that are reflected by the RIS two or more times is assumed to be negligible \cite{Wu2019intelligentJ,  Huang2019Reconfigurable, Guo2019Weighted, Yu2019Enabling, Chen2019Intelligent, Jiang2019Over}, \cite{Ding2019simple, Yang2019Intelligent, Li2019Joint, Zhu2019Power, Mu2019exploiting, Zuo2020Resource}.
Although it is generally difficult to obtain perfect CSI in RIS-aided wireless networks, various channel estimation methods have recently been proposed, including brute-force methods \cite{Nadeem2019Intelligent,Yang2019IntelligentOFDM},
 compressive-sensing based methods \cite{He2019Cascaded, Xia2019Intelligent},  and deep learning\cite{Taha2019Enabling}.
  To characterize the theoretical performance gain achieved by RIS, we assume that the  CSI of all channels  is perfectly known \cite{Wu2019intelligentJ,  Huang2019Reconfigurable, Guo2019Weighted,  Yu2019Enabling, Chen2019Intelligent, Jiang2019Over}, \cite{Ding2019simple, Yang2019Intelligent, Li2019Joint, Zhu2019Power, Mu2019exploiting, Zuo2020Resource}.
The proposed algorithm developed in this paper not only serves as the performance upper bound for the practical scenarios with imperfect CSI, but also provides useful 
 insights on the design of RIS-empowered NOMA schemes. %, which can motivate more research efforts in this area. 

In NOMA systems, all the users sequentially perform  SIC to decode their own signals in a specific order.
It has been shown in \cite{Ding2016UserPairing} that the decoding order  plays  a critical role in  determining the overall system performance.
In   SISO NOMA networks without RIS, the users are usually ordered based on their channel conditions with respect to the BS. 
In MISO NOMA networks without RIS, the user ordering becomes more complicated and is determined by the product of the channel gain and the beamforming gain of each user \cite{Liu2018Multiple}. 
  Furthermore, in RIS-empowered MISO NOMA networks, obtaining the optimal user ordering is further complicated, as the concatenated channel response $(\bm h_{r,k}^{\sf{H}}\bm \Theta\bm G+\bm h_{d,k}^{\sf H})$ depends not only on $\bm h_{r,k}$,  $\bm G$, and $\bm h_{d,k}$, but also on the phase-shift matrix $ \bm \Theta$. 
With the exhaustive search method, there are $K!$ different decoding orders for $K$ users. 
We denote the set of all possible user orderings as $\mathcal{S} =\{ \mathcal{S}_1, \ldots, \mathcal{S}_u, \ldots, \mathcal{S}_{K!}\}$, where $\mathcal{S}_u =\{1_u, \ldots, k_u, \ldots,K_u\}$ denotes the $u$-th user ordering scheme. 
For a given user ordering $\mathcal{S}_u$,  user $U_{k_{u}} $, $k > 1$,   sequentially decodes and removes the signals intended for users $\{U_{1_u}, \ldots, U_{{(k-1)}_u}\}$, whereas the signals of other users are treated as noise. 
  The remaining signal at user $U_{l_u}$,  $l_u \ge k_u$, to detect the signal intended for user $U_{k_u}$
   can be expressed as
\begin{eqnarray}\label{receive2}
y_{l_u}^{k_u} = 
\left(\bm h_{r,l_u}^{\sf{H}}\bm \Theta\bm G+\bm h_{d,l_u}^{\sf H} \right)\sum_{j=k_u}^{K_u}\bm w_js_j+e_{l_u}. 
\end{eqnarray}
According to \eqref{receive2}, the achievable SINR for user $U_{l_u}$ to decode the signal that is intended to user $U_{k_u}$ can be expressed as
\begin{eqnarray}\label{SINR}
\text{SINR}_{l_u}^{k_u} = \frac{|(\bm h_{r,l_u}^{\sf{H}}\bm \Theta\bm G+\bm h_{d,l_u}^{\sf H})\bm w_{k_u}|^2}{\sum_{j=(k+1)_u}^{K_u} |(\bm h_{r,l_u}^{\sf{H}}\bm \Theta\bm G+\bm h_{d,l_u}^{\sf H})\bm w_j |^2+ \sigma^2}, \nonumber \\
k_u \in \mathcal{S}_u, \forall l_u \geq k_u.
\end{eqnarray}
To successfully decode the $k_u$-th signal at user $U_{l_u}$  (i.e., $l_u\geq k_u$), the achievable SINR of decoding the $k_u$-th signal at user $U_{l_u}$ should be greater than some target  SINR  $\gamma_{k_u}$  for user $U_{k_u}$,
which can be mathematically defined as 
\begin{eqnarray}\label{minsnr}
\mathop{\text{min}}_{l_u\in[k_u,K_u]}  \text{SINR}_{l_u}^{k_u} \geq \gamma_{k_u}.
\end{eqnarray}
It is worth noting that \eqref{minsnr} ensures that all users that are ordered behind user $U_{k_u}$ need to successfully decode the signal intended for user $U_{k_u}$. 
The achievable rate $R_{k_u}$ for  user $U_{k_u}$ can be defined as 
\begin{eqnarray}\label{achievable rate}
R_{k_u} = \text{log}_2 \Big(1+ \mathop{\text{min}}_{l_u\in[k_u,K_u]}  \text{SINR}_{l_u}^{k_u} \Big), \forall k_u \in \mathcal{S}_u,
\end{eqnarray}
where the channel bandwidth is normalized to 1. 
Moreover, to ensure successful SIC for user ordering $\mathcal{S}_u$, the following condition should be satisfied \cite{Alavi2018Beamforming}, \cite{Al2019Energy}
\begin{eqnarray}\label{constraint1}
\!\Big|\!\Big(\!\bm h_{r,k_u}^{\sf{H}} \bm \Theta\bm G\!+\!\bm h_{d,k_u}^{\sf H} \!\Big)\!\bm w_{1_u}\!\Big|^2 \!\geq\!\cdots\!\geq\! \Big|\!\Big(\!\bm h_{r,k_u}^{\sf{H}} \bm \Theta\bm G\!+\!\bm h_{d,k_u}^{\sf H} \!\Big)\!\bm w_{k_u}\!\Big|^2  \nonumber \\ 
\!\!\! \geq \!\cdots\! \geq\! \Big|\!\Big(\!\bm h_{r,k_u}^{\sf{H}}\bm \Theta\bm G\!+\!\bm h_{d,k_u}^{\sf H} \!\Big)\!\bm w_{K_u}\!\Big|^2\!\!, k_u \!\in\!\mathcal{S}_u.
\end{eqnarray}
Note that the above inequalities are defined to ensure that the users are ordered in a descending order of their received power.

\subsection{Problem Formulation}
In this subsection, we formulate a total transmit power minimization problem by jointly optimizing the beamforming vectors (i.e., $\{\bm w_k, k\in \mathcal{K}\}$) at the BS and the phase-shift matrix (i.e., $\bm \Theta$) at the RIS, taking into account the data rate requirements of all users and the unit modulus constraints of all reflecting elements.  
For a given decoding order $\mathcal{S}_u$, the optimal total transmit power is denoted as $\sum_{k=1_u}^{K_u}\|\bm w_k^{\star}\|^2$, where $\{\bm w_k^{\star}\}$ denotes  the  solution of the following optimization problem. 
For notational ease, we omit the decoding order index $u$ in the sequel.
The total transmit power minimization problem is formulated as  
\begin{eqnarray}\label{mixed}
\mathscr{P}_1:\mathop{\text{minimize}}_{ \{\bm w_k\},\bm \Theta}
&&\sum_{k=1}^{K}\|\bm w_k\|^2 \nonumber\\
\text{subject to}&& \text{log}_2\left(1+ \mathop{\text{min}}_{l\in[k,K]} \text{SINR}_l^k \right) \geq R_k^{\text{min}},\forall \, k, \label{cons1} \\
\label{eqn_theta}&&|\bm \Theta_{n,n}| = 1, \forall \, n,
\end{eqnarray}
where $\|\bm w_k\|^2$ is the transmit power allocated to user $U_k$ and  
$R_k^{\text{min}}$ denotes the minimum data rate requirement of user $U_k$.   Since the SINRs required to support consecutive SIC are satisfied through the minimum rate constraints $\eqref{cons1}$,  constraints $\eqref{constraint1}$ become unnecessary in Problem $\mathscr{P}_1$ \cite{Li2019Joint}, \cite{Alavi2018Beamforming}, \cite{Zhu2020optimal}. 
The joint  optimization of beamforming vectors and the phase-shift matrix in $\mathscr{P}_1$ is more difficult than the power minimization problem in \cite{Alavi2018Beamforming, Wu2019intelligentJ} due to the complicated non-convex constraints $\eqref{mixed}$ with highly coupled variables and minimization operator. 
%In addition, the SOCP programming in \cite{wu2018intelligentmag} can not be directly applied to solve the $\mathscr{P}_1$

To make the above constraints $\eqref{mixed}$ more tractable, we rewrite  constraints  $\eqref{mixed}$  as
\begin{eqnarray}\label{achievable rate constraint}
\mathop{\text{min}}_{l\in[k,K]} \text{SINR}_l^k \geq \gamma_k^{\text{min}}, \forall \, k,
\end{eqnarray}
where $\gamma_k^{\text{min}} = 2^{R_k^{\text{min}}} -1$ is the minimum SINR required to successfully decode signal $s_k$. 
To eliminate the minimization operator, constraints $\eqref{achievable rate constraint}$ can be further rewritten as
\begin{eqnarray}\label{simplyrate}
 &&\gamma_k^{\text{min}} \left(\sum_{j=k+1}^{K}\big|(\bm h_{r,l}^{\sf{H}}\bm \Theta\bm G+\bm h_{d,l}^{\sf H})\bm w_j\big|^2 + \sigma^2\right)   \leq \nonumber \\
&& \ \ \ \ \ \ \ \ \ \ \ \      |(\bm h_{r,l}^{\sf{H}}\bm \Theta\bm G+\bm h_{d,l}^{\sf H})\bm w_k|^2,
   \, \forall k, l = k,\ldots, K.
\end{eqnarray}
Therefore, Problem $\mathscr{P}_1$ can be equivalently rewritten as
\begin{eqnarray}\label{mixedSINR}
\mathscr{P}_2:\mathop{\text{minimize}}_{ \{ \bm w_k\},\bm \Theta}
&&\sum_{k=1}^{K}\|\bm w_k\|^2\nonumber \\
\text{subject to}&&   \mathrm{constraints}\; \eqref{eqn_theta}, \eqref{simplyrate}. 
\end{eqnarray}

 Problem $\mathscr{P}_2$ is still highly intractable due to the non-convex bi-quadratic constraints  \eqref{simplyrate}, in which the beamforming vectors and the phase-shift matrix are highly coupled, and the non-convex  unit modulus constraints \eqref{eqn_theta}.
To address these unique challenges, we present an  alternating DC optimization framework to solve Problem $\mathscr{P}_2$ in Sections III, IV, and V. 
We present an exhaustive search to obtain the optimal total transmit power $P^{\star}$, i.e., $P^{\star} = \min_{\mathcal{S}_u\in \mathcal{S}} \sum_{k=1_u}^{K_u}\|\bm w_k^{\star}\|^2$ among all possible user orderings.
%The optimal total transmit power $P^{\star}$ can be obtained by exhaustively searching over all possible decoding orders, i.e., $P^{\star} = \min_{\mathcal{S}_u\in \mathcal{S}} \sum_{k=1_u}^{K_u}\|\bm w_k^{\star}\|^2$.
To further reduce the computational complexity, we develop an efficient user ordering scheme and present in a closed-form in Section VI. We will then demonstrate that it can achieve almost the same performance as the exhaustive search method in Section VII. 

\section{Alternating Optimization Framework}
Recently,  most the existing works, e.g.,  \cite{Huang2019Reconfigurable, Wu2019intelligentJ,  Guo2019Weighted}  applied  a powerful alternating optimization to  design the active  beamforming at the BS and the phase-shift matrix at the RIS, which always yields convex constraints  given the phase-shift matrix (i.e., affine constraints \cite{Huang2019Reconfigurable},
second-order cone (SOC) constraints \cite{Wu2019intelligentJ},   and quadratic constraints  \cite{Guo2019Weighted}).
Inspired by this, in this section, we shall employ alternating optimization  to decompose the joint optimization Problem $\mathscr{P}_2$ into two subproblems but with non-convex constraints, i.e., a non-convex  QCQP subproblem for transmit beamforming vectors and a non-convex QCQP feasibility subproblem for the phase-shift matrix.
 \subsection{Transmit Beamforming Vectors Optimization}
For a given phase-shift matrix $ \bm\Theta $,  the concatenated channel response $\bm h_l^{\sf H}= \bm h_{r,l}^{\sf{H}}\bm \Theta\bm G+\bm h_{d,l}^{\sf H} \in \mathbb{ C}^{1\times M}$ is fixed, and hence $ \mathscr{P}_2 $ is simplified as the following non-convex QCQP problem
\begin{eqnarray}\label{fixtheta}
\mathop{\text{minimize}}_{\{ \bm w_k\}}
&&\sum_{k=1}^{K}\|\bm w_k\|^2\nonumber \\
\text{subject to}&&  \gamma_k^{\text{min}} \left(\sum_{j=k+1}^{K}|\bm h_{l}^{\sf{H}}\bm w_j|^2 + \sigma^2\right) \leq |\bm h_{l}^{\sf{H}}\bm w_k|^2,  \nonumber \\
&&\ \ \ \ \ \ \ \ \ \ \ \ \ \  \forall \,k, l = k,\ldots, K.
\end{eqnarray}

While the problem $\eqref{fixtheta}$  appears similar to the beamforming design problem \cite{Wu2019intelligentJ, Bjornson2014optimal}, it cannot be equivalently transformed into a second-order cone program (SOCP) optimization problem because it is impossible to have a phase rotation to simultaneously satisfy $\mathfrak{R} (\bm h_{k}^{\sf{H}}\bm w_k) = \mathfrak{R}(\bm h_{l}^{\sf{H}}\bm w_k) = 0,  k < l \leq K$.
One may exploit the approximation technique  to relax the optimization problem as an SOCP problem, which, however, results in a suboptimal solution with performance degradation\cite{Alavi2018Beamforming}. 
 To further improve the performance,
  the semidefinite relaxation (SDR) technique \cite{Ma2010Semidefinite} can be exploited by formulating the optimization problem as a semidefinite programming (SDP)  form  by lifting $\bm w_k$ into a positive semidefinite (PSD) matrix $\bm W_k \in \mathbb{ C}^{M\times M}$, where $\bm W_k = \bm w_k\bm w_k^{\sf H}$ and
  $\mathrm{rank}(\bm W_k) = 1, \forall k$, followed by dropping rank-one constraints \cite{Alavi2018Beamforming, Zhu2020optimal}.
  If the resulting SDP problem yields rank-one matrices, then it will also be  the optimal solution to the original problem.  
  If not, the Gaussian randomization technique \cite{Ma2010Semidefinite} can be adopted to obtain a suboptimal solution, which however leads to performance degradation.
  Furthermore, the returned solution fails to satisfy the rank-one with high probability when the dimension of optimization variables is high \cite{Chen2017ADMM}. It is also equally important to note that the existing methods fail to accurately detect the feasibility of  multiple rank-one constraints, which may yield performance degradation as the early stopping in the procedure of alternating optimization.

\subsection{Phase-Shift Matrix Optimization}
Given the beamforming vectors $ \{\bm w_k, k \in \mathcal{K}\}$,  we denote $b_{l,k} = \bm h_{d,l}^{\sf H}\bm w_k$  and $ \bm a_{l,k} = \text{diag}(\bm h_{r,l}^{\sf H})\bm G\bm w_k $,  $\forall \, k, l\geq K$. Hence, we have $(\bm h_{r,l}^{\sf{H}}\bm \Theta\bm G+\bm h_{d,l}^{\sf H})\bm w_k = \bm v^{\sf H}\bm a_{l,k} + b_{l,k}$, where $ \bm v=[e^{j\theta_1},\ldots,e^{j\theta_N}] ^{\sf H}$. Thus, Problem $ \mathscr{P}_2 $ is simplified as the following  feasibility detection problem
\begin{eqnarray}\label{fixomega}
\mathop{\text{Find}}
&&{\bm v}\nonumber \\
\text{subject to}&& \gamma_k^{\text{min}} \left(\sum_{j=k+1}^{K}|\bm v^{\sf H} \bm a_{l,j}+ b_{l,j}|^2 + \sigma^2\right) \leq \nonumber \\
&&\ \ \ \ \ \ \ \ \ \ \ \ \ \  |\bm v^{\sf H} \bm a_{l,k}+ b_{l,k}|^2,  \forall \, k, l = k,\ldots, K,\nonumber\\
&&|\bm v_n| =1, \forall \, n = 1,\ldots, N.
\end{eqnarray}

Problem $\eqref{fixomega}$ is  non-convex and inhomogeneous due to the non-convexity of the quadratic constraints and the unit modulus constraints. 
To an efficient algorithm design, it can be reformulated by introducing an auxiliary variable $t$ as follows: 
\begin{eqnarray}\label{fixomega1}
\mathop{\text{Find}}
&&{\tilde{\bm v}}\nonumber \\
\text{subject to}&&  \gamma_k^{\text{min}} \left(\sum_{j=k+1}^{K}\left(\tilde{\bm v}^{\sf H} \bm R_{l,j}\tilde{\bm v} +  |b_{l,j}|^2\right) +\sigma^2\right) \leq  \nonumber \\
&&\ \ \ \ \ \ \ \ \ \ \ \ \ \ \tilde{\bm v}^{\sf H} \bm R_{l,k}\tilde{\bm v}+ |b_{l,k}|^2, \forall \, k, l = k,\ldots, K,\nonumber\\
&&|\tilde{\bm v}_n| =1, \forall \, n = 1,\ldots,N+1,
\end{eqnarray}
where
\begin{eqnarray}
 \bm R_{l,k} = 
\begin{bmatrix}
        \bm a_{l,k}\bm a_{l,k}^{\sf H}     &    \bm a_{l,k}     b_{l,k}^{\sf H}       \\
        b_{l,k}\bm a_{l,k}^{\sf H}       & 0 
\end{bmatrix} \;\; \textrm{and} \quad \tilde{\bm v}=
\begin{bmatrix}
\bm v    \\
t
\end{bmatrix}.
\end{eqnarray}
If we obtain a feasible solution, denoted as $ \tilde{\bm v}^{\star}$, by solving Problem $\eqref{fixomega1}$, then a feasible solution to Problem $\eqref{fixomega}$ can immediately be obtained by setting $\bm{v}^{\star} =   [\tilde{\bm v}^{\star}/\tilde{\bm v}^{\star}_{N+1}]_{(1:N)}$, where $[\bm x]_{(1:N)}$ denotes the first $N$ elements of vector $\bm x$. 
 A natural method  is  to  formulate the Problem  $\eqref{fixomega}$ as an SDP optimization problem by lifting $\bm \tilde{\bm v} $ into a PSD matrix $\bm V \in \mathbb{ C}^{N+1 \times N+1}$, where $\bm V = \tilde{\bm v}\tilde{\bm v}^{\sf H}$ and $\mathrm{rank}(\bm V) = 1$, following by dropping the rank-one constraint via the SDR technique, similar to\cite{Wu2019intelligentJ}. 
 However,  such an SDR technique often yields a solution which fails to satisfy the rank-one constraint, especially in the case where $N$ becomes very large\cite{Chen2017ADMM}. 
 Moreover, the suboptimal solution obtained by the Gaussian randomization technique may not guarantee meeting the original quadratic constraints in $\eqref{fixomega1}$. 
In this case, the procedure of alternating optimization stops early and the performance of NOMA will significantly deteriorate.

Based on the above discussions and different from previous
works \cite{Huang2019Reconfigurable, Wu2019intelligentJ,  Guo2019Weighted}, here we endeavour to address the following coupled challenges to solve the total transmit power minimization problem for RIS-empowered multi-user MISO NOMA networks:
\begin{itemize}
        \item For the transmit beamforming vectors optimization, we need to efficiently solve a non-convex QCQP optimization problem with the capability of  accurately detecting the feasibility of multiple rank-one constraints in the lifted matrix space to minimize the total transmit power, thereby achieving the performance gains;
        \item For the phase-shift matrix optimization, we need to accurately detect the feasibility of a non-convex quadratically constrained problem, which can avoid the early stopping of the alternating optimization procedure, thereby enhancing the performance compared with state-of-the-art methods.
\end{itemize}
In summary, we shall develop an efficient method to  accurately detect the feasibility of the non-convex quadratic constraints to push the alternating optimization procedure for the transmit beamforming vectors design and phase-shift matrix design alternatively.
In the next section, to address the limitations of existing  methods,
 we present a unified DC method to solve the aforementioned subproblems.

\section{A Unified Difference-of-Convex Programming Method}
In this section, we propose a unified DC method  to the non-convex QCQP subproblems to accurately detect the feasibility of non-convex quadratic constraints for  the transmit beamforming vectors optimization and  the phase-shift matrix optimization while avoiding the alternative procedure to stop early, thereby achieving the performance improvement.
The basic idea is to reformulate the resultant non-convex QCQP subproblems in each iteration as multiple rank-constrained matrices optimization problems via matrix lifting, following by equivalently representing the rank-one constraint as the difference between the nuclear norm and the spectral norm. We  first present an exact DC representation for  the rank-one constraint and then reformulate  the resultant non-convex QCQP subproblems into DC programs.
\subsection{DC Representation for Rank-One Constraint}
Firstly, we present an exact DC representation for the rank-one constraint.
For a matrix $\bm X\in \mathbb{C}^{N\times N}$, the rank-one constraint can be rewritten as
\begin{eqnarray}\label{equrank}
\big\| [\sigma_1(\bm X),\ldots,\sigma_i(\bm X) \ldots,\sigma_N(\bm X)]\big\|_0 = 1,
\end{eqnarray}
where $\sigma_i(\bm X)$ is the $i$-th largest singular value of matrix $\bm X$, and $\|\cdot\|_0$  is the $l_0$-norm of a vector. It is noteworthy that the rank function  is a discontinuous
function.
%, i.e., the number of non-zero entries. 
%Note that the nuclear norm and the spectral norm are, respectively, given by
%\begin{eqnarray}\label{twonorm}
%\|\bm X\|_* = \sum_{i=1}^{N} \sigma_i(\bm X) \ \ \textrm{and} \ \ \|\bm X\|_2 = \sigma_1(\bm X).
%\end{eqnarray}
To reformulate a continuous function, we introduce an exact DC representation for the rank-one constraint in Proposition 1.
\begin{proposition} 
        For a PSD matrix $\bm X\in \mathbb{C}^{N\times N}$ with $\text{Tr}(\bm X)>0$, we have  $\mathrm{rank}(\bm X)=1\Leftrightarrow \|\bm X\|_*-\|\bm X\|_2=0$.
\end{proposition}
\noindent {\it{Proof}}. If matrix $\bm X$ is a rank-one PSD matrix, then the nuclear norm is equal to the spectral norm since $\sigma_i(\bm X)=0$ for all $i\geq 2$. 
Hence, $\|\bm X\|_*-\|\bm X\|_2=\sum_{i=2}^{N} \sigma_i(\bm X)=0$ implies that $\big\| [\sigma_1(\bm X),\ldots,\sigma_N(\bm X)]\big\|_0 \leq 1$.
Because of  $\text{Tr}(\bm X)>0$, we have $\sigma_1(\bm X) >0$. Therefore, $\mathrm{rank}(\bm X)=1$ is equivalent to $ \|\bm X\|_*-\|\bm X\|_2=0$.  $\hfill\square$ 

% whereas the DC representation (i.e., $\|\bm X\|_*-\|\bm X\|_2$) is a continuous function. 
\subsection{Proposed Unified DC Method }
The main idea of our proposed  DC method is to first  reformulate  Problems $\eqref{fixtheta}$ and $\eqref{fixomega1}$ into matrices optimization problems via the matrix lifting technique,  
then apply the DC presentation to accurately detect the rank-one constraints.

Given the phase-shift matrix $\bm \Theta$, by lifting $\bm w_k$ into a PSD matrix $\bm W_k \in \mathbb{ C}^{M\times M}$, where $\bm W_k = \bm w_k\bm w_k^{\sf H}$ and
$\mathrm{rank}(\bm W_k) = 1, \forall k$, 
we reformulate Problem \eqref{fixtheta}  into the following DC program to obtain $K$ rank-one matrices
\begin{eqnarray}\label{liftomegaDC}
\mathscr{P}_3:\mathop{\text{minimize}}_{ \{ \bm W_k\}}
&&\sum_{k=1}^{K}\text{Tr}({ \bm W}_k) + \rho\sum_{k=1}^{K}\Big(\|\bm W_k\|_*-\|\bm W_k\|_2\Big)\nonumber \\
\text{subject to}&& \gamma_k^{\text{min}}\!\! \left(\sum_{j=k\!+\!1}^{K}\!\text{Tr}({\bm H}_l^\mathsf{H}\bm W_j) \!+\! \sigma^2\!\right) \!\leq \!\text{Tr}(\bm H_l^\mathsf{H}\bm W_k), \nonumber\\
 &&\ \ \ \ \ \ \ \ \ \ \ \ \ \ \ \ \ \ \forall \, k, l = k,\ldots, K, \label{omegaDC2}\\
&&\bm W_k \succcurlyeq 0, \forall \, k, \label{omegaDC3}
\end{eqnarray}
where $\rho>0$ is a penalty parameter and  $\bm H_l = \bm h_l \bm h_l^\mathsf{H}\in \mathbb{ C}^{M\times M}$. 
By enforcing the penalty term to be zero, Problem $\mathscr{P}_3$ induces $K$ rank-one matrices. 
After solving $\mathscr{P}_3$, we can recover the beamforming vectors $\bm w_k$ for Problem  \eqref{fixtheta} through the Cholesky decomposition, i.e., $\bm W_k^{\star} = \bm w_k \bm w_k^{\sf H}$, $\forall \, k \in \mathcal{K}$, where $ \{\bm W_k^{\star}, k\in \mathcal{K}\}$ denotes the  solution of Problem $\mathscr{P}_3$.

Similarly, given the beamforming vectors $\{\bm w_k, k \in \mathcal{K} \}$, we  minimize the  difference between the nuclear norm and the spectral norm by  lifting $\bm \tilde{\bm v} $ into a PSD matrix $\bm V \in \mathbb{ C}^{N+1 \times N+1}$, where $\bm V = \tilde{\bm v}\tilde{\bm v}^{\sf H}$, $\mathrm{rank}(\bm V) = 1$. That is,
\begin{eqnarray}\label{liftvDC}
\!\!\!\!\!\!\mathscr{P}_4: \mathop{\text{minimize}}_{\bm V}
&&\|\bm V\|_*-\|\bm V\|_2\nonumber \\
\text{subject to}&&  \gamma_k^{\text{min}}\!\!\left (\sum_{j=k+1}^{K}\left(\text{Tr}(\bm R_{l,j}\bm V) \!+\! |b_{l,j}|^2\right)\!+\! \sigma^2\!\right) \!\leq\!   \nonumber\\
&&\ \ \ \ \text{Tr}(\bm R_{l,k}\bm V) \!+\!|b_{l,k}|^2, \forall k,  l\! = \!k,\!\ldots,\! K, \label{vDC1}\\
&&\bm V_{n,n} =1, \forall \, n = 1,\ldots,N+1, \label{vDC2}\\
&&\bm V \succcurlyeq 0. \label{vDC3}
\end{eqnarray}
Specifically, when the objective value of Problem $\mathscr{P}_4$ becomes zero, we  obtain an exact rank-one feasible solution, denoted as $\bm V^{\star}$. 
Using Cholesky decomposition $\bm V^{\star} = \tilde{\bm v}  \tilde{\bm v}^{\sf H}$, we then obtain a feasible solution $\tilde{\bm v} $ to Problem $\eqref{fixomega1}$. 

Although the above DC programs are still non-convex, they have the algorithmic advantage. 
 In the next section, we will develop the  DC algorithm  for problems $\mathscr{P}_3$ and $\mathscr{P}_4$ via successive convex relaxation.  The superior performance has been shown in vast applications, e.g., degrees-of-freedom maximization for data shuffling in wireless distributed computing \cite{Yang2019Data} and model aggregation via over-the-air computation for federated learning \cite{Yang2020Federated}.
 We further prove that the sequence solutions of the algorithm converges to the stationary point. 
 The proposed alternating DC method outperforms the alternating SDR method, which will be demonstrated through numerical simulations in Section VII.

\section{Alternating DC Algorithm with  Convergence Guarantee}
In this section,  we shall propose an efficient  DC algorithm  to obtain high-quality solutions for the beamforming vectors and the phase-shift matrix  with convergence  guarantee. 
By rewriting the objective function as the difference of two strongly convex functions, we further prove that the presented algorithm  converges to the stationary solution in each iteration.  Furthermore, the  alternating DC method always effectively converges.

\subsection{Difference of  Strongly Convex Functions Representation }
Although the DC programs $\mathscr{P}_3$ and $\mathscr{P}_4$ are still non-convex, they have a good structure that can be exploited to develop an efficient algorithm, which successively solves the convex relaxation  of the primal problem and its dual problem \cite{Dinh1997d.c}. 
In order to establish some important properties of the  algorithm, we represent the DC objective function as the difference of two strongly convex functions. Specifically, we  rewrite $\mathscr{P}_3$ as
\begin{eqnarray}\label{DC1}
\mathop{\text{minimize}}_{ \{ \bm W_k\}}
&&f_1 =\sum_{k=1}^{K}\text{Tr}({ \bm W}_k) + \rho\sum_{k=1}^{K}\Big(\|\bm W_k\|_*-\|\bm W_k\|_2\Big) + \nonumber \\
 &&\ \ \ \ \ \ \ \ \ \ \ \ \ \ \mathit{I}_{\mathcal{C}_1}(\{\bm W_k\}),
\end{eqnarray}
and Problem $\mathscr{P}_4$ as
\begin{eqnarray}\label{DC2}
\mathop{\text{minimize}}_{\bm V}
&&f_2=\|\bm V\|_*-\|\bm V\|_2 + \mathit{I}_{\mathcal{C}_2}(\bm{V}),
\end{eqnarray}
where $\mathcal{C}_1$ and $\mathcal{C}_2$  denote the PSD cones that  satisfy the constraints in problems $\mathscr{P}_3$ and $\mathscr{P}_4$, respectively, and the indicator function is defined as
$$\mathit{I}_{\mathcal{C}}(\bm{X})=
\begin{cases}
0,& \bm{X}\in \mathcal{C},\\
+\infty,& \text{otherwise}.
\end{cases}$$

We rewrite the DC functions $f_1$ and $f_2$ as the difference of two strongly convex functions, i.e., $f_1 = g_1-h_1$ and $f_2 = g_2-h_2$, where
\begin{eqnarray}
&&g_1= \sum_{k=1}^{K}\text{Tr}(\bm W_k) + \rho\sum_{k=1}^{K}\|\bm W_k\|_* +\mathit{I}_{\mathcal{C}_1}(\{\bm W_k\}) + \nonumber \\
&&\ \ \ \ \ \ \ \ \ \ \ \ \ \  \ \ \   \frac{\eta}{2} \sum_{k=1}^{K} \|\bm W_k\|_F^2,\\
&& h_1 = \rho\sum_{k=1}^{K}\|\bm W_k\|_2 +  \frac{\eta}{2} \sum_{k=1}^{K} \|\bm W_k\|_F^2, \\
&& g_2 = \|\bm V\|_*+ \mathit{I}_{\mathcal{C}_2}(\bm{V}) + \frac{\eta}{2} \|\bm V\|_F^2,\\
&& h_2 = \|\bm V\|_2 + \frac{\eta}{2} \|\bm V\|_F^2.
\end{eqnarray}
Because of the additional quadratic terms (i.e., $\frac{\eta}{2} \sum_{k=1}^{K} \|\bm W_k\|_F^2$ and $\frac{\eta}{2} \|\bm V\|_F^2$), $g_1$, $h_1$, $g_2$, and $h_2$ are all $\eta$-strongly convex functions. It turns out that problems $\eqref{DC1}$ and $\eqref{DC2}$ have the unified structure of minimizing the difference of two strongly convex functions, i.e., 
\begin{eqnarray}\label{DC structure}
\mathop{\textrm{minimize}}_{ \bm{Z}\in\mathbb{ C}^{ m\times m} } &&f_i = g_i(\bm{Z}) - h_i(\bm{Z}),  i = 1, 2.
\end{eqnarray}  

To solve the non-convex DC program, we present a  DC algorithm  to construct a sequence of candidates to the primal and dual solutions via successive convex relaxations      
in the sequel. 

\subsection{ DC Algorithm for Problem $\eqref{DC structure}$ }
 According to the Fenchel's duality \cite{Rockafellar2015Convex}, the dual problem of Problem $\eqref{DC structure}$ is equivalent to
\begin{eqnarray}\label{dual DC structure}
\mathop{\textrm{minimize}}_{ \bm{Y}\in\mathbb{ C}^{m\times m}} &&h_i^*(\bm{Y}) - g_i^*(\bm{Y}), i = 1, 2, 
\end{eqnarray}  
where   $g_i^*$ and $h_i^*$ are the conjugate functions of $g_i$ and $h_i$, respectively. 
The conjugate function $h_i^*(\bm{Y})$ is defined as
\begin{eqnarray}\label{conjugate func}
\!\!h_i^*(\bm{Y}) = \mathop{\textrm{sup}}_{\bm{Y}\in \mathbb{ C}^{m\times m}}  \{ \langle \bm{Z}, \bm{Y}\rangle - h_i(\bm{Z}) : \bm Z \in \mathcal{Z} \}, i = 1, 2, 
\end{eqnarray}  
where the inner product is defined as  $\langle \bm{X}, \bm{Y}\rangle = \mathfrak{R}(\text{Tr}(\bm{X}^\mathsf{H}\bm{Y}))$ according to Wirtinger calculus \cite{Dong2018demixing} in the complex domain and $\mathcal{Z}$ denotes $\bm{Z}$'s  feasible solution region.
 Since the primal problem $\eqref{DC structure}$ and its dual problem $ \eqref{dual DC structure}$ are still non-convex, the DC algorithm iteratively updates both the primal and dual variables via successive convex approximations. 
Specifically, in the $r$-th iteration, we have 
\begin{eqnarray}
\!\!\!\!\!\!\!\!\!\!\!&&\bm{Y}^r = \mathop{\textrm{arg\,inf}}_{ \bm{Y}}\  h_i^*(\bm{Y})-\big[g_i^*(\bm{Y}^{r-1}) + \langle \bm{Y}-\bm{Y}^{r-1}, \bm{Z}^r\rangle\big],\\ \label{r-iteration}
\!\!\!\!\!\!\!\!\!\!\!&&\bm{Z}^{r+1} = \mathop{\textrm{arg\,inf}}_{ \bm{Z}}\  g_i(\bm{Z})-\big[h_i(\bm{Z}^{r}) + \langle \bm{Z}-\bm{Z}^{r}, \bm{Y}^r\rangle\big]. \label{r-iterationprimal}
\end{eqnarray}     
Based on the Fenchel bi-conjugation theorem \cite{Rockafellar2015Convex}, 
the solution to Problem $\eqref{r-iteration}$ can be written as $\bm{Y}^r  \in \partial_{\bm{Z}^r}h_i$, %\cite{Yang2019Data, Yang2020Federated},    
where $\partial_{\bm{Z}^r}h_i$ is the sub-gradient of $h_i$ with respect to $\bm{Z}$ at $\bm{Z}^r$. 
Thus, $\{\bm W_k^{r}, k\in \mathcal{K}\}$ at the $r$-th iteration for $\mathscr{P}_3$ can be obtained by solving the following convex  problem 
\begin{eqnarray}\label{subliftomegaDC}
\mathop{\text{minimize}}_{ \{ \bm W_k\}}
&&g_1-\sum_{k=1}^{K}\langle \bm{W_k},\partial_{\bm{W_k}^{r-1}}h_1\rangle\nonumber \\
\text{subject to}&&  \mathrm{constraints}\; \eqref{omegaDC2},\eqref{omegaDC3}.
\end{eqnarray}
Similarly, $\bm V^r$ at the $r$-th iteration for  $\mathscr{P}_4$ can be obtained by solving the following convex optimization problem
\begin{eqnarray}\label{subliftvDC}
\mathop{\text{minimize}}_{\bm V}
&&g_2-\langle \bm{V},\partial_{\bm{V}^{r-1}}h_2\rangle\nonumber \\
\text{subject to}&& \mathrm{constraints}\; \eqref{vDC1}, \eqref{vDC2}, \eqref{vDC3}.
\end{eqnarray}

Problems $\eqref{subliftomegaDC}$ and $\eqref{subliftvDC}$ are convex and can be efficiently solved  by using CVX \cite{grant2014cvx}. Note that $\partial_{\bm{W_k}^{r-1}}h_1 $ and $\partial_{\bm{V}^{r-1}}h_2$ are, respectively, given by
\begin{eqnarray}
&&\partial_{\bm{W_k}^{r-1}}h_1 = \rho\partial_{\bm{W_k}^{r-1}}\|\bm W_k\|_2 + \eta \bm W_k^{r-1}, \\
&&\partial_{\bm{V}^{r-1}}h_2 = \partial_{\bm{V}^{r-1}}\|\bm V\|_2 + \eta \bm V^{r-1}.
\end{eqnarray}
It is worth noting that the sub-gradient of  $\|\bm X\|_2$ at $ \bm{X}^{r}\in \mathbb{C}^{N\times N}$ (i.e., $\bm\partial_{\bm{X}^{r}}\|\bm{X}\|_2$) can be efficiently computed according to the following proposition. 
\begin{proposition}
 For a PSD matrix $\bm X$, the sub-gradient of $\|\bm X\|_2$ at $ \bm{X}^{r}$ can be efficiently computed as $\bm u_1 \bm u_1^{\sf H}$, where $\bm u_1\in\mathbb{ C}^{N}$ is the eigenvector corresponding to the largest eigenvalue $\sigma_1( \bm{X}^r)$. 
\end{proposition}
The  efficient DC algorithm  is developed by successively solving the convex relaxation of the primal and dual problems of DC programming. 
The overall algorithm, solving problems $\mathscr{P}_3$ and $\mathscr{P}_4$ in an alternative approach, which is referred to as the alternating DC algorithm as presented in Algorithm $\ref{algo2}$. 
Specifically, Algorithm 1 optimizes $\{\bm w_k, k\in \mathcal{K}\}$ and $\bm \Theta$ alternatively, where the  presented DC algorithm is adopted to obtain the beamforming vectors and the phase shifts in the lifted matrix  space that satisfy the rank-one constraints.
 For a fair comparison, the alternating DC algorithm terminates when  the decrease of the objective value of Problem $\mathscr{P}_2$ is smaller than  $\epsilon$, which is a predetermined convergence threshold, or Problem $\eqref {fixomega}$ becomes infeasible.
We shall prove the convergence of Algorithm $\ref{algo2}$ in the sequel.
\begin{algorithm}[h]
	\SetKwData{Left}{left}\SetKwData{This}{this}\SetKwData{Up}{up}
	\SetKwInOut{Input}{Input}\SetKwInOut{Output}{output}
	\Input{ $\bm \Theta^0$ and threshold $\epsilon>0$.}
	\For{$t=1,2,\ldots$}{
		Given  $\bm \Theta^{t-1}$, solve Problem $\mathscr{P}_3$  and obtain the  solution $\{\bm W_k^{t}, k\in \mathcal{K}\}$.\\
		\For{$r =1,2,\ldots$}{
			Select  a subgradient $\partial_{\bm{W_k}^{r-1}} \| \bm{W_k}\|_2, k\in \mathcal{K}$. \\
			Solve Problem $\eqref{subliftomegaDC}$ and obtain the  solution $\{\bm W_k^r, k\in \mathcal{K}\}$.\\
			\If{ penalty component of Problem $\mathscr{P}_3$ is zero}{\textbf{break}}}
		Obtain $\{\bm w_k^{t}, k\in \mathcal{K}\}$ via Cholesky decomposition $\bm W_k^{r} = \bm w_k^{t}   (\bm w_k^{t})^{\sf H}$. \\
		
		Given  $\{\bm w_k^{t}, k\in \mathcal{K}\}$,  solve Problem $\mathscr{P}_4$  and obtain a feasible solution $\bm V^{t}$.\\
		\For{$r =1,2,\ldots$}{
			Select  a subgradient $\partial_{\bm{V}^{r-1}} \|  \bm{V}\|_2$. \\
			Solve  Problem $\eqref{subliftvDC}$  and obtain the  solution $\bm V^r$.\\
			\If{ objective value of  Problem $\mathscr{P}_4$  is  zero}{\textbf{break}}} 
		Obtain $\tilde{\bm v}^{t}$ via Cholesky decomposition $\bm V^{r} = \tilde{\bm v}^{t}(\tilde{  \bm v}^t)^{\sf H}$ and $\bm v^{t} =   [\tilde{\bm v}^{t}/ \tilde{\bm v}_{N+1}^{t}]_{(1:N)} $.\\  
		
		\If{ decrease of the total transmit power is below $\epsilon$ or Problem $\eqref {fixomega}$ becomes infeasible}{\textbf{break}}}
	\caption{Proposed Alternating DC Algorithm for Solving Problem  $\mathscr{P}_2$.}
	\label{algo2}
\end{algorithm}
\subsection{Alternating DC Algorithm Convergence Analysis}
Before proving the convergence of the proposed alternating DC algorithm, 
we present some important properties of the solutions obtained by solving the convex relaxation of the primal and dual problems of DC programming in the following proposition. 
\begin{proposition}  
For any $r =0, 1,\ldots,$ the sequence $\{\bm W_k^r, k\in \mathcal{K}\}  $ generated by iteratively solving Problem $\eqref{subliftomegaDC}$
 has the following properties:\\
\indent(i) The sequence $\{\bm W_k^r, k\in \mathcal{K}\}  $ converges to a stationary point of $f_1$ in Problem $\eqref{DC1}$ from an arbitrary initial point, and the sequence  $\{f_1^r \}$ is strictly decreasing and convergent.\\
\indent(ii) For any $r =0, 1,\ldots,$ we have 
\begin{eqnarray}\label{covergence}
\text{Avg}\Big(\big\|\bm W_k^{r}-\bm W_k^{r+1} \big\|\Big) \leq \frac{f_1^{0}-f_1^{\star}}{\eta(r+1)},\forall \, k = 1,\ldots, K,
\end{eqnarray}
where $f_1^{\star}$ is the global minimum of $f_1$ and $\text{Avg}\Big(\big\|\bm W_k^{r}-\bm W_k^{r+1} \big\|\Big)$ denotes the average of the sequence $\{ \|\bm W_k^i - \bm W_k^{i+1}\|_F^2  \}_{i=0}^r$.

Likewise, for any $r =0, 1,\ldots$, the sequence $\{ \bm V^r \}$ generated by iteratively solving Problem $\eqref{subliftvDC}$
 has the following properties:\\
\indent(iii) The sequence $\{ \bm V^r  \}$ converges to a stationary point of $f_2$ in Problem $\eqref{DC2}$ from an arbitrary initial point, and the sequence of $\{f_2^r \}$ is strictly decreasing and convergent.

(iv) For any $r = 0, 1,\ldots,$ we have 
\begin{eqnarray}\label{covergence1}
\text{Avg}\Big(\big\|\bm V^{r}-\bm V^{r+1} \big\|\Big) \leq \frac{f_2^{0}-f_2^{\star}}{\eta(r+1)},
\end{eqnarray}
where $f_2^{\star}$ is the global minimum of $f_2$.
\end{proposition}
\noindent {\it{Proof}}. Please refer to Appendix A. $\hfill\square$

Based on Proposition 3, the convergence analysis of Algorithm $\ref{algo2}$ is given in  proposition 4.
\begin{proposition}  
The objective value of Problem $\mathscr{P}_2$ in $\eqref{mixedSINR}$ decreases as the number of iteration increases until convergence by applying the proposed alternating DC algorithm.
\end{proposition}
\noindent {\it{Proof}}.  Please refer to Appendix B. $\hfill\square$
\section{  Low-Complexity User Ordering Scheme}
Solving Problem $\mathscr{P}_2$ $K!$ times by exhaustive search to obtain the optimal user ordering is computationally prohibitive when $K$ is large. To address this issue, we shall develop a low-complexity user ordering scheme to determine the decoding order of the users  for RIS-empowered MISO NOMA networks. The existing studies mainly focused on either the channel condition \cite{Ding2016UserPairing} or QoS \cite{Zhou2018Stable} based user ordering scheme for NOMA transmission. 
Different from the existing studies, we order the users according to the minimum transmit power required to meet their data rate requirements. 
It is worth noting that the proposed user ordering criterion captures the effects of both the channel conditions and the QoS requirements.
Specifically, the minimum transmit power required at the BS to serve user $U_k$  can be obtained by solving the following problem
\begin{eqnarray}\label{original order}
\mathop{\text{minimize}}_{\bm w_k, \bm \Theta}&&\|\bm w_k\|^2\nonumber\\
\text{subject to}&& \|(\bm h_{r,k}^{\sf{H}}\bm \Theta\bm G+\bm h_{d,k}^{\sf H})\bm w_k\|^2\label{obj}\geq \gamma_k^{\text{min}}\sigma^2,\nonumber\\
&&|\bm \Theta_{n,n}| = 1, \forall \, n.
\end{eqnarray}
Although \cite{Wu2019intelligentJ} has provided a SDR-based alternating optimization approach to solve $\eqref{original order}$,  this user ordering scheme needs to solve $K$ SDP  problems and suffers from a very high computation complexity.
To reduce this computation complexity, we propose a method  to derive solutions for $\bm w_k$ and $\bm \Theta$ with closed-form expressions in this following.

 For a given $\bm \Theta$, it is well-known that the maximum-ratio transmission (MRT) is the optimal transmit beamforming solution to Problem \eqref{original order} \cite{Wu2019intelligentJ}, i.e., $\bm w_k^{\star} = \sqrt{p_k}\frac{(\bm h_{r,k}^{\sf{H}}\bm \Theta\bm G+\bm h_{d,k}^{\sf H})^{\sf H}}{\|\bm h_{r,k}^{\sf{H}}\bm \Theta\bm G+\bm h_{d,k}^{\sf H}\|}$,
where  $p_k$ is the transmit power of the AP for user $U_k$.  Furthermore, the optimal transmit power  $p_k^{\star}$ satisfies $p_k^{\star}  = \frac{\gamma_k^{\text{min}}\sigma^2}{\|\bm h_{r,k}^{\sf{H}}\bm \Theta\bm G+\bm h_{d,k}^{\sf H}\|^2}$, which is determined by the target data rate and combined channel gains.
Therefore, minimizing the transmit power is equivalent to maximizing the combined channel power gain, which is given by
\begin{eqnarray}\label{givenomega1}
\mathop{\text{maximize}}_{\bm \Theta}
&& \|\bm h_{r,k}^{\sf{H}}\bm \Theta\bm G+\bm h_{d,k}^{\sf H}\|^2\nonumber\\
\text{subject to}&&|\bm \Theta_{n,n}| = 1, \forall \, n.
\end{eqnarray}

Similar to Problem $\eqref{fixomega}$, by introducing $\tilde{  \bm v}$,  Problem $\eqref{givenomega1}$ can be rewritten as 
\begin{eqnarray} \label{addauxi}
\mathop{\text{maximize}}_{ \tilde{  \bm v}\in \mathbb{ C}^{N+1}}
&& \tilde{  \bm v}^{\sf H}\bm Q_k \bm \tilde{  \bm v} \nonumber \\
\text{subject to}&&|\tilde{  \bm v}|_n = 1, \forall n = 1,\ldots,N+1,
\end{eqnarray}
where 
\begin{eqnarray}
\bm Q_k = 
\begin{bmatrix}
\text{diag}(\bm h_{r,k}^{\sf H})\bm G \bm G^{\sf H}\text{diag}(\bm h_{r,k})    &   \text{diag}(\bm h_{r,k}^{\sf H})\bm G \bm h_{d,k}    \\
\bm h_{d,k}^{\sf H}\bm G^{\sf H} \text{diag}(\bm h_{r,k})        & \bm h_{d,k}^{\sf H}\bm h_{d,k}
\end{bmatrix}.
\end{eqnarray}
Problem $\eqref{addauxi}$ has  a concave objective function with non-convex unit modulus constraints. 
The authors in \cite{Wu2019intelligentJ} and \cite{Yang2019Intelligent} applied the SDR technique to reformulate Problem $\eqref{addauxi}$ into a convex SDP problem. To further reduce the computational complexity, we relax the unit modulus constraints in Problem \eqref{addauxi} as a norm constraint, i.e., $\|\tilde{ \bm v}\|^2 = N + 1$.  Such a relaxation yields a closed-form expression of $\bm \Theta$ for  Problem \eqref{addauxi}. 
The relaxed optimization problem is given by
\begin{eqnarray} \label{addauxirelaxed}
\mathop{\text{maximize}}_{ \tilde{  \bm v}\in \mathbb{ C}^{N+1}}
&& \tilde{  \bm v}^{\sf H}\bm Q_k \bm \tilde{  \bm v} \nonumber \\
\text{subject to}&&\|\tilde{  \bm v}\|^2= N+1.
\end{eqnarray}

Problem \eqref{addauxirelaxed} is  an eigenvalue problem and its optimal solution is given by
\begin{eqnarray} \label{solution}
\tilde{  \bm v}^{\star}= \sqrt{N+1} \bm u_1,
\end{eqnarray}
where $\bm u_1 \in \mathbb{C}^{N+1}$ is the eigenvector corresponding to the largest eigenvalue $\sigma_1(\bm Q_k)$ of matrix $\bm Q_k$.  Thus, the optimal objective value of Problem $\eqref{addauxirelaxed}$ is $\sigma_1(\bm Q_k)(N+1)$.  The objective value of Problem $\eqref{original order}$ is approximated by $\hat{p}_k = \frac{\gamma_k^{\text{min}}\sigma^2}{\sigma_1(\bm Q_k)(N+1)}$, which depends on  the target data rate and the channel condition. 
We order $K$ users in the descending order of power $\hat{p}_k$.
Specifically, the user with the largest value of $\hat{p}_k$ decodes its own signal first, while the user with the smallest value of $\hat{p}_k$ needs to decode all other users' signals before decoding its own signal. 

By deriving the ordering criterion  in a closed form, the proposed user ordering scheme enjoys a much low computational complexity. 
Numerical results in the next section will show that the proposed user ordering scheme
achieves almost the same performance as the SDR-based user ordering scheme, which needs to solve $K$ SDP  problems and suffers from a much higher computation complexity.   
Furthermore, the proposed user ordering scheme only suffers slight performance degradation compared to the exhaustive search scheme.

\section{Numerical Results}

In this section, we present sample numerical results to demonstrate the effectiveness of the proposed alternating DC method in RIS-empowered downlink NOMA networks. 
We consider a three-dimensional coordinate system, where the BS is located at $(0, 0, 10)$ meters. 
The RIS is placed at $(50, 50, 15)$ meters. 
In addition, the users are  randomly distributed in the region of $(-50,50,0)\times(60,160,0)$ meters. 
The path loss model under consideration is $L(d) = T_0{d}^{-\alpha}$,
where $T_0=-30$ dB is the path loss at  reference distance one meter, $d$ is the link distance, and $\alpha$ is the path loss exponent. 
The path loss exponents for the BS-user link, the BS-RIS link, and the RIS-user link are set to 3.5, 2, and 2.2, respectively \cite{Wu2019intelligentJ, Jiang2019Over, Wu2019BeamformingDiscrete }.  Each antenna at the AP is assumed to have an isotropic radiation pattern with 0 dBi antenna gain, while each reflecting element of RIS is assumed to have 3 dBi gain, since each RIS reflects signals only in its front half-space \cite{Griffin2009Complete}.
All channels are assumed to suffer from Rayleigh fading \cite{Huang2019Reconfigurable, Jiang2019Over}. 
We denote $d_{\mathrm{BU}}^k$, $d_{\mathrm{IU}}^k$, and $d_{\mathrm{IB}}$ as the distances between user $U_k$ and the BS, between user $U_k$ and the RIS, and  between the BS and the RIS, respectively. 
Hence, the corresponding channel coefficients are given by $\bm h_{d,k} = \sqrt{L(d_{\mathrm{BU}}^k)}\bm\gamma^d$,
$\bm h_{r,k} = \sqrt{L(d_{\mathrm{IU}}^k)}\bm\gamma^r$,
$\bm G= \sqrt{L(d_{\mathrm{IB}})}\bm\Gamma$,
where $\bm \gamma^d\sim \mathcal{CN}(0,\bm I)$, $\bm \gamma^r\sim \mathcal{CN}(0,\bm I)$,  and $\bm \Gamma\sim \mathcal{CN}(0,\bm I)$. Unless specified otherwise,  we set $R_k^{\text{min}}=1.5$ bits per channel use, $\forall \, k \in \mathcal{K}$, $\sigma^2 =-80$ dBm, $\rho = 10$, and $\epsilon = 1e-4$.
All results in Figs. \ref{fig:iter}-\ref{fig:discretephase} are obtained by averaging over 100 channel realizations.
 \subsection{ Performance Comparison of Different Methods}
 We compare the proposed alternating DC method with  two state-of-the-art methods:
 \begin{itemize}
 \item  \textbf{Alternating SDR}: This method leverages the SDR technique to  solve problems $\eqref{fixtheta}$ and $\eqref{fixomega1}$ alternatively. The Gaussian randomization technique is further applied when the solution obtained by the SDR method does not satisfy the rank-one constraints. 
 \item  \textbf{Random phase shift}: With this method, the phase-shift matrix $\bm \Theta$ is randomly chosen and kept fixed when solving the transmit power minimization problem $\mathscr{P}_3$. 
 \end{itemize}

\begin{figure}[t]
	\centering
	\begin{minipage}{.46\textwidth}
		\centering
		\includegraphics[width=7.5cm,height=6cm]{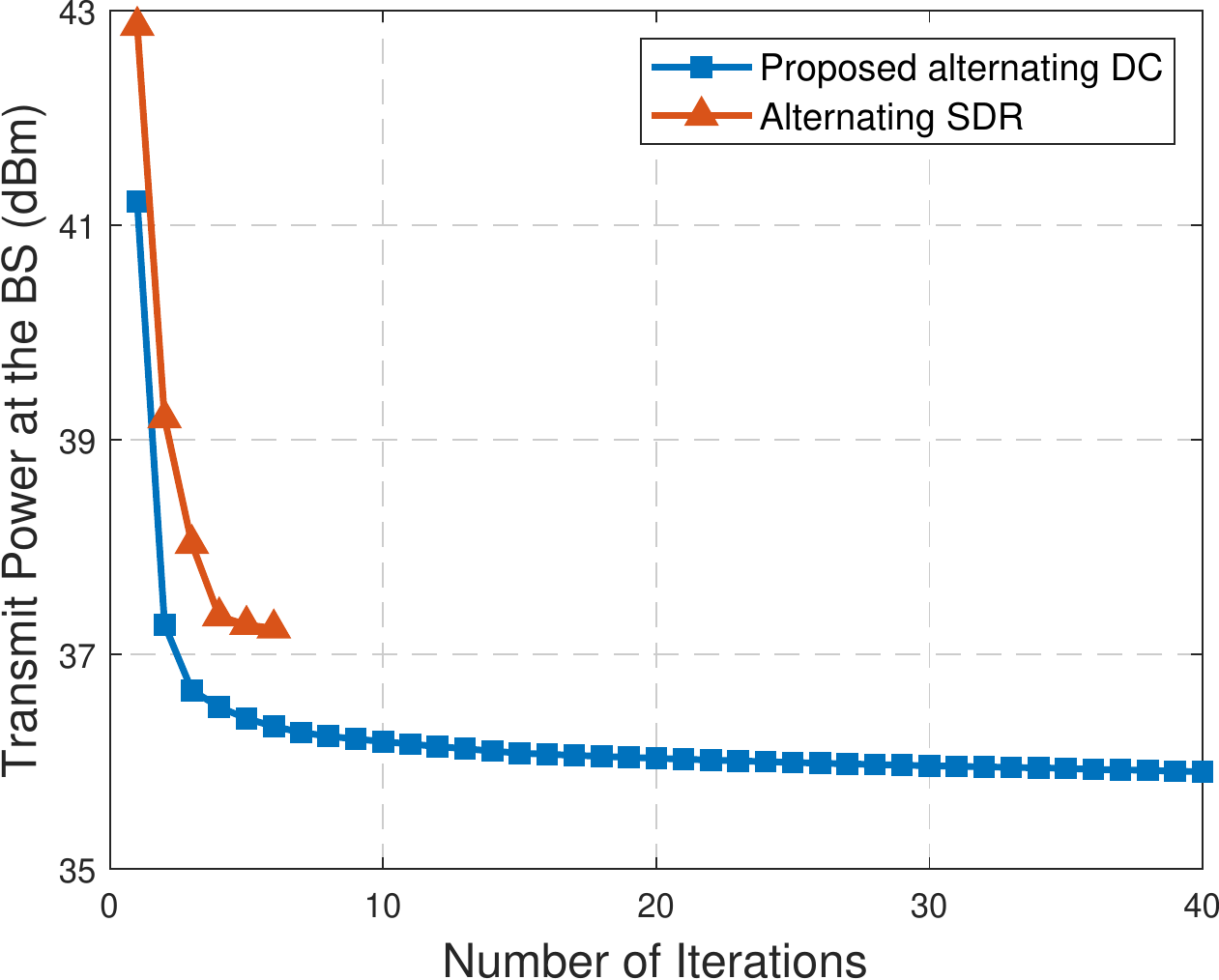}
		\vspace{-3mm}
		\caption{Convergence behaviors of the proposed alternating DC method and the alternating SDR method. }\label{fig:iter}
		\vspace{4mm}
	\end{minipage}
	\hspace{4mm}
	\begin{minipage}{.46\textwidth}
		\centering
		\includegraphics[width=7.5cm,height=6cm]{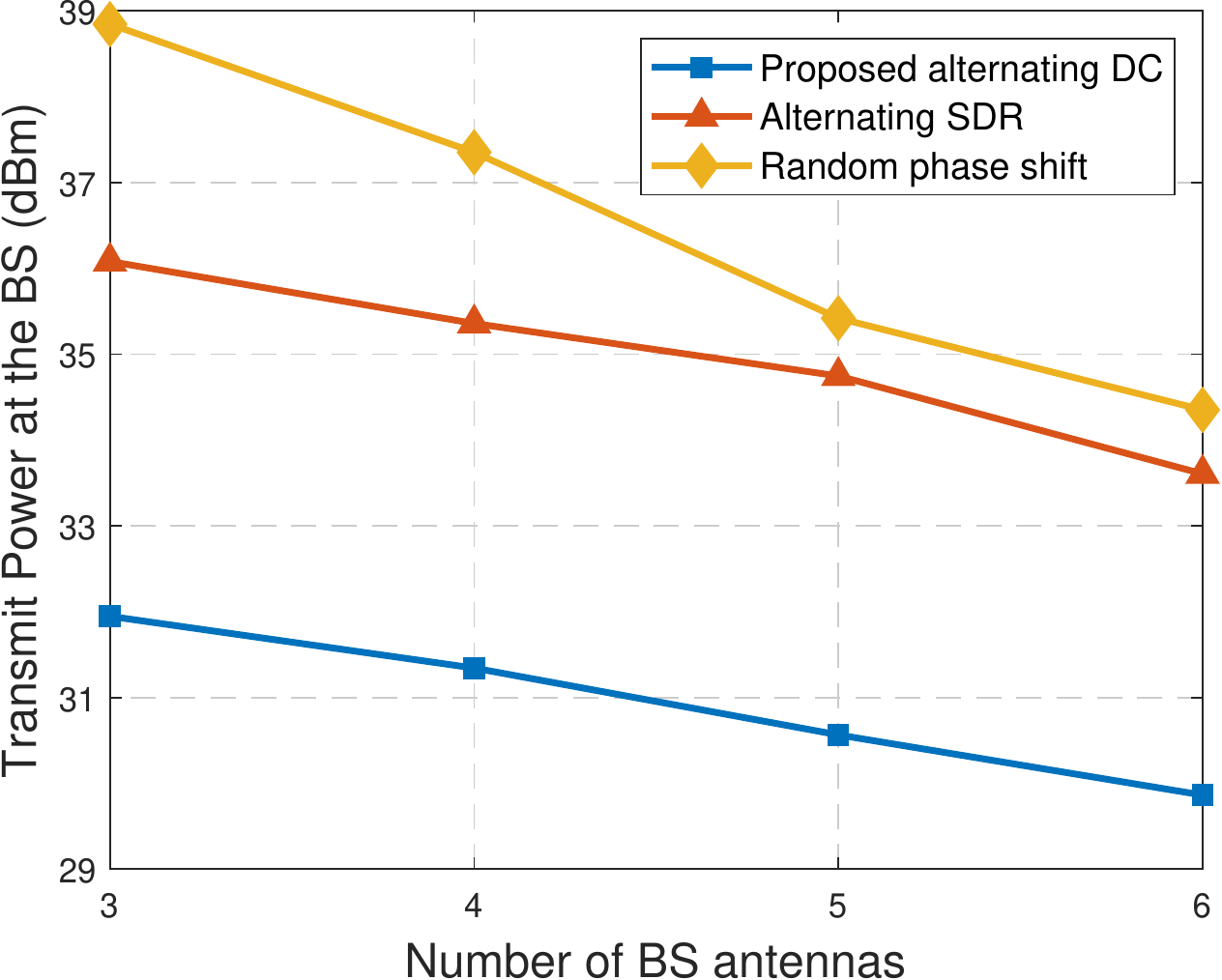}
		\vspace{-3mm}
		\caption{Transmit power versus the number of  BS antennas when $N=20$ and $K=7$.}\label{fig:M}
		\vspace{-1mm}
	\end{minipage}
\end{figure}

Fig. \ref{fig:iter} illustrates the convergence behaviors of the proposed alternating DC method and the alternating SDR method   when $K = 8$, $M = 7$, and $N = 20$. 
The  transmit power obtained by the alternating SDR method is higher than that obtained by the alternating DC method at the first iteration when the same initial phase-shift matrix is given. This is  because removing  the rank-one constraints incurs performance degradation, while the proposed alternating DC method ensures that the rank-one constraints hold.
Furthermore, it  can be observed that the alternating SDR method with Gaussian randomization fails to return a feasible solution to Problem $\eqref{fixomega}$ after the sixth iteration. 
In contrast, the proposed alternating DC method is able to induce exact rank-one  solutions, and hence accurately detects the feasibility of Problem $\eqref{fixomega}$, which avoids the early stopping in the alternating optimization procedure, thereby considerably reducing the transmit power consumption compared with alternating SDR method.

Fig. \ref{fig:M} shows the impact of the number of BS antennas (i.e., $M$) on the total transmit power when $N=20$ and $K=7$.  
The total transmit power of the BS decreases as the value of $M$ increases, which indicates that more antennas at the BS lead to a better performance by achieving a higher diversity gain.
In addition, both the proposed alternating DC method and the alternating SDR method significantly outperform the random phase shift method. 
Hence demonstrating the necessity of jointly optimizing the beamforming vectors at the BS and the phase-shift matrix at the RIS. 
Furthermore, due to the superiority of the proposed DC representation, the proposed alternating DC method consumes much less transmit power than the alternating SDR method.
 
\begin{figure}[t]
	\centering
	\begin{minipage}{.46\textwidth}
		\centering
		\includegraphics[width=7.5cm,height=6cm]{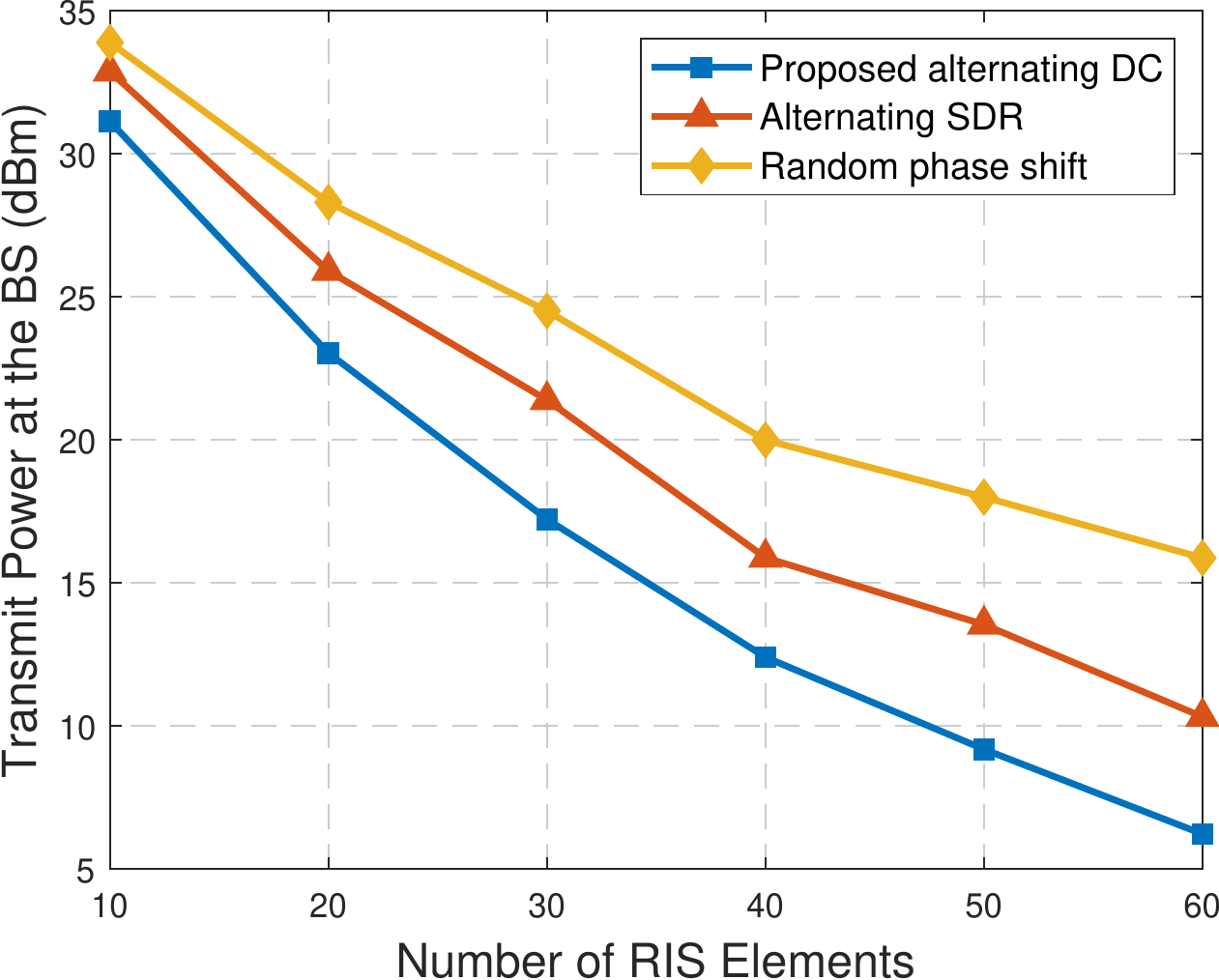}
		\vspace{-3mm}
		\caption{Transmit power  versus  the number of RIS elements when $M=4$ and $K=5$.}\label{fig:N}
		\vspace{4mm}
	\end{minipage}
	\hspace{4mm}
	\begin{minipage}{.46\textwidth}
		\centering
		\includegraphics[width=7.5cm,height=6cm]{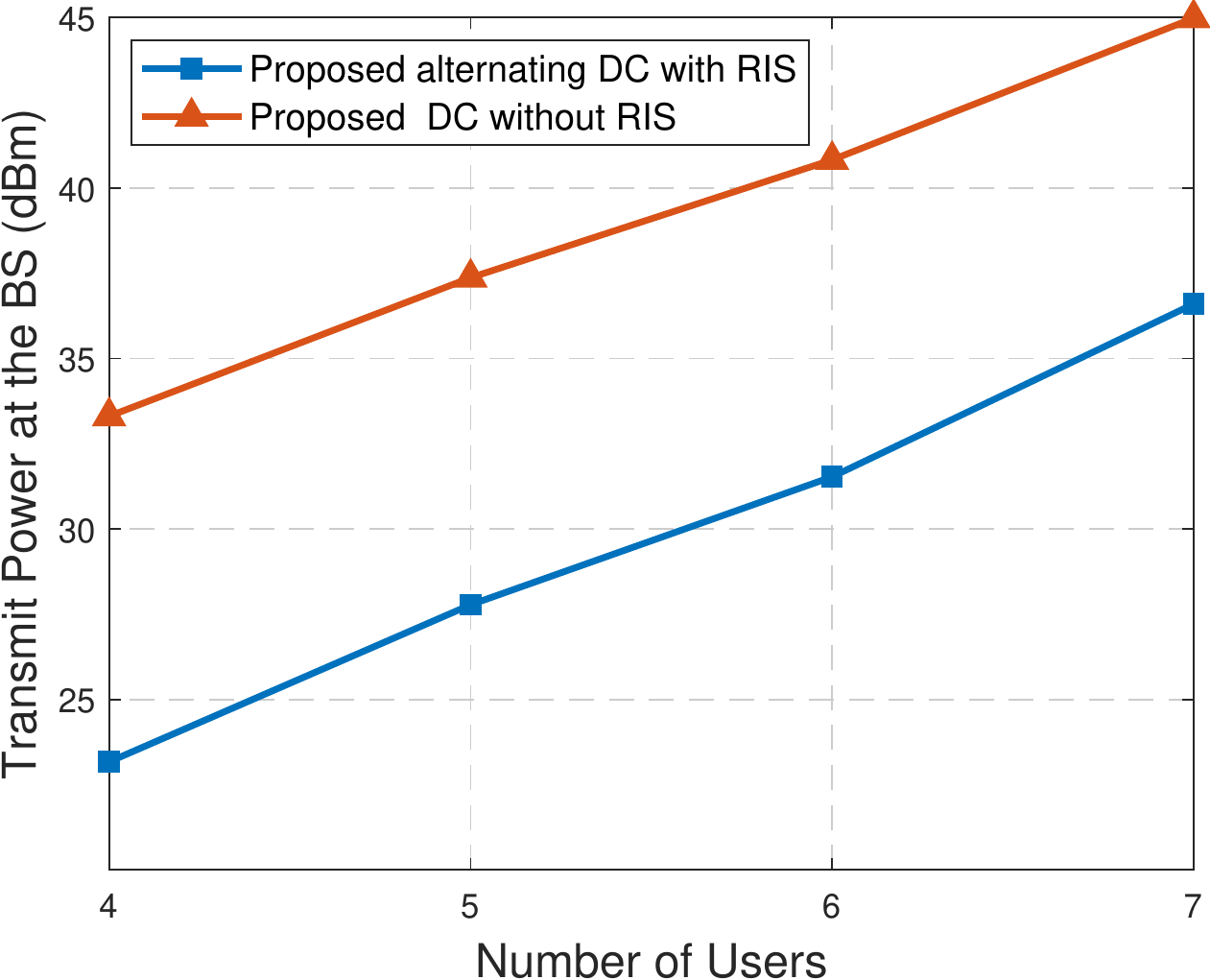}
		\vspace{-3mm}
		\caption{Transmit power  versus the number of users for downlink NOMA networks with and without RIS.}\label{fig:K}
		\vspace{-1mm}
	\end{minipage}
\end{figure}

Fig. \ref{fig:N} illustrates the impact of the number of passive reflecting elements at the RIS (i.e., $N$) on the total transmit power when $M=4$ and $K =5$. 
For all the three methods under consideration, the total transmit power decreases quickly as the value of $N$ increases. 
This is because an RIS with more reflecting elements significantly enhances the receiving power at the users and introduces more channel differences among the users.
Therefore, a larger number of passive reflecting elements leads to a higher energy-efficiency. 

Fig. \ref{fig:K} shows the performance of downlink NOMA networks with and without RIS when $M=3$ and $N =15$. 
The transmit power of NOMA networks without RIS is obtained by solving Problem $\mathscr{P}_3$ by setting $\bm \Theta = \bm 0$, where the users are ordered in an ascending order of the channel conditions of the direct links.
The RIS-empowered network outperforms the network without RIS even when the number of passive reflecting elements is small, which demonstrates the effectiveness of deploying RIS in cellular networks. 

%\begin{figure}[t]
%       \centering
%       \includegraphics[width=8.5cm,height=6.5cm]{Kloaded.pdf} 
%       \vspace{-0.4cm}
%       \caption{Transmit power  versus number of users for downlink NOMA networks with and without RIS.}\label{fig:K}
%\end{figure}

\subsection{Performance Comparison of Different  User Ordering Schemes}
We compare the performance of the proposed user ordering scheme with that of other benchmarks as are listed below:
\begin{itemize}
\item \textbf{Direct link quality based user ordering}: In this user ordering scheme, the users are ordered according to the quality of the BS-user link, e.g., $\|\bm h_{d,k}\|_2$.
\item \textbf{Exhaustive search based user ordering}: This user ordering scheme finds the optimal decoding order that achieves the best performance by exhaustively searching over all $K!$ possible decoding orders.
\item \textbf{Proposed user ordering}: 
The proposed user ordering scheme with closed-form solutions is presented in Section VI, where the users are ordered according to the qualities of the BS-user link, the BS-RIS link, and the RIS-user link, as well as, the target data rate. 
\item \textbf{SDR-based user ordering}: 
Different from the proposed scheme, this user ordering scheme obtains the ordering criterion by solving Problem $\eqref{givenomega1}$ using the SDR technique \cite{Yang2019Intelligent}. 
\end{itemize}
\begin{figure}[t]
	\centering
	\begin{minipage}{.46\textwidth}
		\centering
		\includegraphics[width=7.5cm,height=6cm]{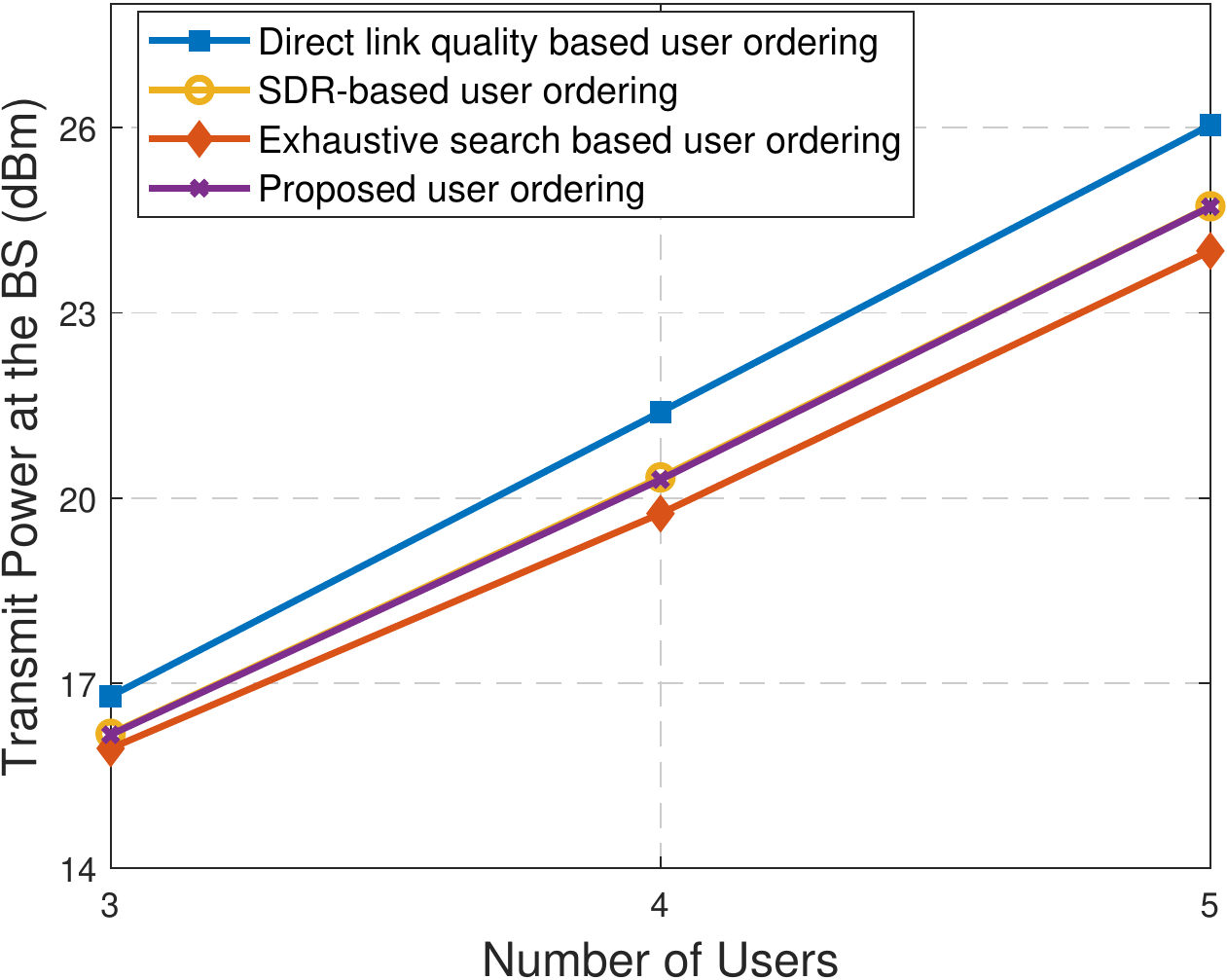}
		\vspace{-3mm}
		\caption{Transmit power versus the number of users for different user ordering schemes.}\label{fig:order}
		\vspace{4mm}
	\end{minipage}
	\hspace{4mm}
	\begin{minipage}{.46\textwidth}
		\centering
		\includegraphics[width=7.5cm,height=6cm]{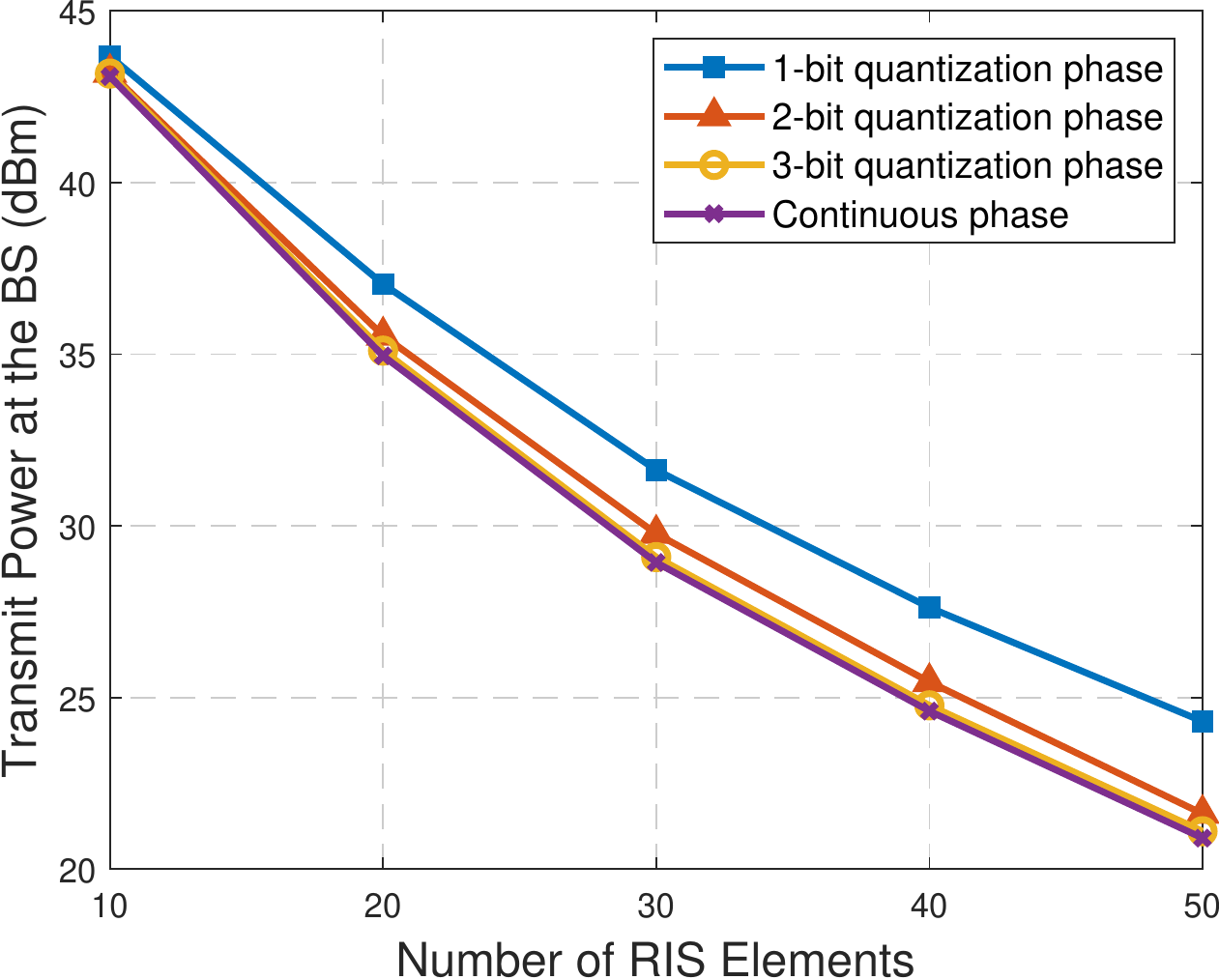} 
		\vspace{-3mm}
		\caption{Transmit power versus the number of RIS elements with discrete  and continuous phase shifts.}\label{fig:discretephase}
		\vspace{-1mm}
	\end{minipage}
\end{figure}
%\begin{figure}[t]
%       \centering
%   \includegraphics[width=8.5cm,height=6.5cm]{KorderLoaded.pdf} 
%   \vspace{-0.4cm}
%\caption{Transmit power versus number of users for different user ordering schemes.}\label{fig:order}
%\end{figure}

Fig. \ref{fig:order} compares the performance of our proposed user ordering scheme with three benchmarks when $N = 20$ and $M =2$. 
It is observed that the performance gaps between the optimal user ordering scheme and other three user ordering schemes increase as the number of users increases. 
However, the optimal user ordering scheme needs to exhaustively search all $K!$ possible decoding orders, and its computational complexity is extremely high.   
Although the direct link quality based user ordering scheme has the lowest complexity, it suffers from a larger performance degradation than the proposed and the SDR-based user ordering schemes. This is because the last two schemes capture the effects of users' target data rates and combined channels which have an impact on the transmit power at BS for the each user.  
When $K=5$, the transmit powers of the direct link quality based, the proposed, and the SDR-based user ordering schemes are $2.04$ dBm, $0.72$ dBm, and $0.72$ dBm higher than the optimal user ordering scheme, respectively. 
Moreover, as shown in Fig. \ref{fig:order}, the proposed user ordering scheme achieves almost the same performance as the SDR-based user ordering scheme. 
However, the SDR-based user ordering scheme needs to solve $K$  SDP problems  for $K$ users, the complexity of which is much higher than the proposed user ordering scheme with closed-form solutions. 

\subsection{Performance Comparison of Discrete and Continuous Phase Shifts}
In practical systems, the RIS with a large number of passive elements has a finite phase resolution, which depends on the number of quantization bits,  denoted as $B$ \cite{Wu2019BeamformingDiscrete}. We numerically investigate the effect of RIS's finite phase resolution on the total transmit power consumption  at the BS. 
In simulations, each optimized continuous phase shift $ \theta_n$ is quantized to its nearest discrete value in the set $\Big\{0, \frac{2\pi}{2^{B}},\ldots, \frac{2\pi\times (2^B-1)}{2^{B}}  \Big\}$.  

 Fig. \ref{fig:discretephase}   plots the total transmit power consumption of the proposed alternating DC method versus the number of RIS elements for different phase quantization bits  when $K = 6$, $M =5$, and $R_k^{\text{min}}=2$  bits per channel use. 
We observe that the total transmit power consumption of the network with a discrete phase-shift RIS is greater than that of the network with a continuous phase-shift RIS. 
As the value of $B$ increases, the total transmit power consumption decreases. 
With 1 or 2-bit phase shifters, the performance gap between the continuous and discrete phase shifts increases as the number of passive elements increases. 
Moreover, 3-bit phase shifters is practically sufficient to achieve almost the same performance as the continuous phase shifters. 
 
\section{Conclusions}
In this paper, we studied an overloaded RIS-empowered NOMA network to minimize the total transmit power by jointly optimizing the beamforming vectors at the BS and phase-shift matrix at RIS, where the RIS  is capable of inducing desirable channel differences among the users to enhance the performance of NOMA.
To address the unique challenges of highly coupled optimization variables and non-convex quadratic constraints, we proposed an alternating DC method to solve the non-convex bi-quadratically constrained quadratic problem. 
This is achieved by introducing an exact DC representation for the rank-one constraints in the lifted non-convex QCQP problems to accurately detect the feasibility of non-convex quadratic constraints for the transmit beamforming vectors and  phase-shift matrix design. 
Therefore, early stopping can be avoided in the procedure of alternating optimization, thereby considerably improving the performance.
We  also developed an  efficient  DC algorithm with convergence guarantee to solve the resulting DC programming problems via successive convex relaxation. We further proposed a low-complexity user ordering scheme, which achieves a comparable performance to the exhaustive search method.
Simulation results demonstrated that the proposed alternating DC method outperforms the state-of-the-art methods in terms of total transmit power minimization.

This initial investigation demonstrated the effectiveness of deploying RIS in NOMA networks for energy efficiency enhancement. Numerical results also showed that RIS with low phase resolution can achieve almost the same performance as RIS with continuous phase shifts. More works are needed to fully exploit the benefits of RIS-empowered NOMA systems, including theoretical analysis, channel estimation, large-scale optimization.
For future studies, the resource allocation framework developed in this paper will be extended to the scenario with multiple BSs, while taking into account channel estimation for practical implementations.  

\section*{Appendix} 

\subsection{ Proof of Proposition 3 }  \label{app_prop3}
         Without loss of generality, we shall only present the proofs of properties (i) and (ii), while properties (iii) and (iv) can be proved similarly. 
        
         We first present the proof of property (i). For the sequence $ \{  \bm W_k^r, k\in \mathcal{K} \}  $ generated by iteratively solving problem $\eqref{subliftomegaDC}$, we denote the dual variables as $  \bm Y_k^r \in \partial_{\bm{W_k}^{r}}h_1 $. Due to the strong convexity of $h_1$, we have
         \begin{eqnarray}
        &&  h_1^{r+1} \geq h_1^{r} +  \sum_{k=1}^{K} \langle \Delta_r \bm W_k,\bm Y_k^r\rangle+ \frac{\eta}{2} \sum_{k=1}^{K}\|\Delta_r \bm W_k\|_F^2,\label{strongh11}\\
           &&\sum_{k=1}^{K} \langle \bm W_k^r,\bm Y_k^r\rangle = h_1^r + (h_1^*){^r},\label{strongh12}
         \end{eqnarray}
     where $ \Delta_r \bm W_k = \bm W_k^{r+1}-\bm W_k^r$. By adding $g_1^{r+1}$ to both sides of        $\eqref{strongh11}$, we obtain 
     \begin{eqnarray}\label{addg1}
     \!\!\!\!\!\!\!\!\!\!\!\!\!\!&&f_1^{r+1} \leq g_1^{r+1}\!-\!h_1^{r}\!-\!\sum_{k=1}^{K} \langle \Delta_r \bm W_k,\bm Y_k^r\rangle\!-\!\frac{\eta}{2} \sum_{k=1}^{K}\|\Delta_r \bm W_k\|_F^2, 
     \end{eqnarray}
%    where $f_1^{r+1} = g_1^{r+1} - h_1^{r+1}$.
      
    For the update of primal variable $\{\bm W_k, k\in \mathcal{K}\}$ according to $\eqref{r-iterationprimal}$, we have $ \bm Y_k^r \in \partial_{\bm{W_k}^{r+1}}g_1 $, which implies that
    \vspace{-4mm} 
    \begin{eqnarray}
     \!\!\!\!\!\!\!\!\!\!\!\!\!&&g_1^{r}  \geq g_1^{r+1} +  \sum_{k=1}^{K} \langle -\Delta_r \bm W_k,\bm Y_k^r\rangle+ \frac{\eta}{2} \sum_{k=1}^{K}\|\Delta_r \bm W_k\|_F^2,\label{strongg11}\\
     \!\!\!\!\!\!\!\!\!\!\!\!\! &&\sum_{k=1}^{K} \langle \bm W_k^{r+1},\bm Y_k^r\rangle =g_1^{r+1} + (g_1^*){^r}.\label{strongg12}
    \end{eqnarray}
    
    Similarly, by subtracting $h_1^r$ from both sides of  $\eqref{strongg11}$, we have
     \begin{eqnarray}\label{add-h1}
     \!\!\!\!\!\!\!\!\!\!\!&&f_1^{r} \geq g_1^{r+1}\!-\!h_1^{r} \!+\!  \sum_{k=1}^{K} \langle -\Delta_r \bm W_k,\bm Y_k^r\rangle\!+\! \frac{\eta}{2} \sum_{k=1}^{K}\|\Delta_r \bm W_k\|_F^2.
    \end{eqnarray}
    
    By subtracting $\eqref{strongg12}$ from  $\eqref{strongh12}$, we have
    \begin{eqnarray}\label{deduce1}
    (f_1^*){^r} \!=\! (h_1^*){^r} \!-\! (g_1^*){^r} \!= \! g_1^{r+1} -h_1^r \!+ \!\sum_{k=1}^{K} \langle -\Delta_r \bm W_k,\bm Y_k^r\rangle,
    \end{eqnarray}
%     where $(f_1^*){^r} = (h_1^*){^r} - (g_1^*){^r}$.  
     
After combining  $\eqref{addg1}$ and $\eqref{deduce1}$, we have
      \begin{eqnarray}\label{deduce2}
     && (f_1^*){^r} \geq f_1^{r+1} + \frac{\eta}{2}\sum_{k=1}^{K}\|\Delta_r \bm W_k\|_F^2.
     \end{eqnarray}
     
     Similarly, after combining  $\eqref{add-h1}$ and $\eqref{deduce1}$, we have 
     \begin{eqnarray}\label{deduce3}
     && f_1^{r} \geq (f_1^*){^r} + \frac{\eta}{2}\sum_{k=1}^{K}\|\Delta_r \bm W_k\|_F^2.
     \end{eqnarray}
     
     Based on $\eqref{deduce2}$ and $\eqref{deduce3}$, we  conclude that
      \begin{eqnarray}\label{decreaseprf}
     && f_1^{r} \geq f_1^{r+1} + \eta \sum_{k=1}^{K}\|\Delta_r \bm W_k\|_F^2.
     \end{eqnarray}
     Therefore, the sequence $\{f_1^r\}$ is non-increasing. Since $f_1 \geq 0$ always holds, we conclude that the sequence $\{f_1^r\}$  is strictly decreasing until convergence, i.e., $\lim_{r\rightarrow\infty} \sum_{k=1}^{K}\|\Delta_r \bm W_k\|_F^2 = 0$.
 
     When the sequence $\{f_1^r\}$ converges at the limit point $\big(\{\bm W_k, k\in \mathcal{K}\}\big)$,  for every limit point, the distances between $\bm W_k^{r+1}$ and $\bm W_k^{r}$ satisfy $\|\bm W_k^{r+1} - \bm W_k^{r}\|_F^2 = 0, \forall \, k = 1,\ldots, K$.
%      \begin{eqnarray}\label{limit point1}
%     &&\|\bm W_k^{r+1} - \bm W_k^{r}\|_F^2 = 0, \forall \, k = 1,\ldots, K.
%     \end{eqnarray}
      Based on inequalities $\eqref{deduce3}$ and $\eqref{decreaseprf}$, the equalities $f_1^{r+1} = f_1^{r} = (f_1^*){^r}$ hold.
%     \begin{eqnarray}
%     && f_1^{r+1} = f_1^{r} = (f_1^*){^r}.
%     \end{eqnarray}
     
       Based on the definitions of $f_1$ and $f_1^*$, we have $ f_1^{r+1} =g_1^{r+1}-h_1^{r+1} $   and  $ (f_1^*){^r} = (h_1^*){^r} - (g_1^*){^r}$.     Therefore, it follows that  
       \vspace{-4mm}
      \begin{eqnarray}\label{limit point2}
      (h_1^*){^r}+ h_1^{r+1} = (g_1^*){^r} + g_1^{r+1}.
     \end{eqnarray}
     
     By combining $\eqref{strongg12}$ and $\eqref{limit point2}$,  we  obtain $ (h_1^*){^r} + h_1^{r+1} = \sum_{k=1}^{K} \langle \bm W_k^{r+1},\bm Y_k^r\rangle.$
%      \begin{eqnarray}\label{limit point3}
%     (h_1^*){^r} + h_1^{r+1} = \sum_{k=1}^{K} \langle \bm W_k^{r+1},\bm Y_k^r\rangle.
%     \end{eqnarray}
     Because $h_1$ is a closed and convex function, we have  $\bm Y_k^r \in \partial_{\bm{W_k}^{r+1}}h_1, \forall k = 1,\ldots, K$.
%     \begin{eqnarray}\label{partial}
%      \bm Y_k^r \in \partial_{\bm{W_k}^{r+1}}h_1, \forall k = 1,\ldots, K.
%     \end{eqnarray}
     Therefore, we have
      \begin{eqnarray}
     \bm Y_k^r \in \partial_{\bm{W_k}^{r+1}}g_1\cap\partial_{\bm{W_k}^{r+1}}h_1, \forall k = 1,\ldots, K.
     \end{eqnarray}
    It is concluded that $\big(\{\bm W_k^{r+1}\}\big)$ is a critical point of $f_1 = g_1 -h_1.$
    
We now present the proof of property (ii). 
Based on the above analysis, we have 
    \begin{eqnarray} \label{Eqn_70}
    \text{Avg}\Big(\sum_{k=1}^K\big\|\bm W_k^{r}-\bm W_k^{r+1} \big\|_F^2\Big) &\leq& \sum_{i=0}^r\frac{1}{\eta(r+1)}(f_1^{i}-f_1^{i-1}) \nonumber \\
     &\leq&\frac{1}{\eta(r+1)}(f_1^{0}-f_1^{r+1}).
    \end{eqnarray}  
    We denote the optimal value of $f_1$ as $f_1^{\star}$. Since inequality $f_1^{\star}\leq f_1^{r+1}$ holds, we have
     \begin{eqnarray} \label{Eqn_71}
      \frac{1}{\eta(r+1)}(f_1^{0}-f_1^{r+1})\leq \frac{1}{\eta(r+1)}(f_1^{0}-f_1^{\star}).
    \end{eqnarray}  
    
According to (\ref{Eqn_70}) and (\ref{Eqn_71}), we conclude that property (ii) holds, i.e.,
     \begin{eqnarray}
    \text{Avg}\Big(\big\|\bm W_k^{r}-\bm W_k^{r+1} \big\|_F^2\Big) \leq \frac{f_1^{0}-f_1^{\star}}{\eta(r+1)},\forall \, k = 1,\ldots,K. 
    \end{eqnarray}
\subsection{Proof of Proposition 4}    
We denote  $f\big(\{\bm w_k\}, \bm \Theta \big)$  as the objective value of $\mathscr{P}_2$ for a feasible solution  $\big(\{\bm w_k\}, \bm \Theta \big)$.
We denote $\big( \{\bm w_k^t\}, \bm \Theta^{t}\big)$ as a  feasible solution of   $\mathscr{P}_2$ at the $t$-th iteration.
For a given $ \bm \Theta^{t}$, we apply the presented  DC algorithm to obtain a solution $\{ \bm W_k^r\}$ for problem $\mathscr{P}_3$, based on which we obtain $\big(\{\bm w_k^t(\bm w_k^{t})^{\sf H}\}\big)$ as the initial point for the $(t+1)$ iteration.  Because the  DC algorithm can accurately detect the feasibility of rank-one constraints, the solution $\{\bm w_k^{t+1}  \}$ can be obtained via cholesky decomposition, where $\bm W_k^{r} = \bm w_k^{t+1}  (\bm w_k^{t+1})^{\sf H}$. Hence, we have $f_1\big(\{ \bm w_k^{t+1}(\bm w_k^{t+1})^{\sf H}\}, \bm \Theta^{t}\big)=f\big(\{ \bm w_k^{t+1}\}, \bm \Theta^{t}\big)$ and $f_1\big(\{ \bm w_k^{t}(\bm w_k^{t})^{\sf H}\}, \bm \Theta^{t}\big)=  f\big(\{ \bm w_k^{t}\}, \bm \Theta^{t}\big)$. 
According to Proposition 3,  the object value of problem $
\mathscr{P}_3$ is strictly decreasing over the iterations. Hence, we have
\begin{eqnarray}
f_1\big(\{ \bm w_k^{t+1}(\bm w_k^{t+1})^{\sf H}\}, \bm \Theta^{t}\big) < f_1\big(\{ \bm w_k^{t}(\bm w_k^{t})^{\sf H}\}, \bm \Theta^{t}\big). 
\end{eqnarray}
Based on  Algorithm $\ref{algo2}$, we have
\begin{eqnarray}\label{allover covergence2}
f\big(\{ \bm w_k^{t+1}\}, \bm \Theta^{t}\big) < f\big(\{ \bm w_k^{t}\}, \bm \Theta^{t}\big).
\end{eqnarray}

For a given $\{ \bm w_k^{t+1}, k\in \mathcal{K}\}$, we also apply the duality-based DC algorithm to solve problem $\mathscr{P}_4$. Based on  Algorithm $\ref{algo2}$, if there exists a feasible solution $\bm V^{t+1}$ to problem $\mathscr{P}_4$, it is also feasible to problem $\eqref{fixomega}$, i.e., $\big(\{ \bm w_k^{t+1} \}, \bm \Theta^{t+1}\big)$ exists. It  follows that 
\begin{eqnarray}\label{allover covergence1}
f\big(\{ \bm w_k^{t+1}\}, \bm \Theta^{t}\big) = f\big(\{ \bm w_k^{t+1}\}, \bm \Theta^{t+1}\big), 
\end{eqnarray}
where the equality holds as the value of $f$ is independent of $\bm \Theta$ but only  depends on $\{ \bm w_k, k\in \mathcal{K}\}$. Based on $\eqref{allover covergence2}$ and $\eqref{allover covergence1}$, we further have 
\begin{eqnarray}\label{allover covergence3}
f\big(\{ \bm w_k^{t+1}\}, \bm \Theta^{t+1}\big) < f\big(\{ \bm w_k^{t}\}, \bm \Theta^{t}\big).
\end{eqnarray}
According to $\eqref{allover covergence3}$, the objective value of problem $\mathscr{P}_2$ is always decreasing over iterations. Therefore, the proposed alternating DC algorithm converges. This completes the proof.

%    \end{appendix} 
  
\bibliographystyle{IEEEtran}
\bibliography{refs}

\begin{thebibliography}{10}
\providecommand{\url}[1]{#1}
\csname url@samestyle\endcsname
\providecommand{\newblock}{\relax}
\providecommand{\bibinfo}[2]{#2}
\providecommand{\BIBentrySTDinterwordspacing}{\spaceskip=0pt\relax}
\providecommand{\BIBentryALTinterwordstretchfactor}{4}
\providecommand{\BIBentryALTinterwordspacing}{\spaceskip=\fontdimen2\font plus
\BIBentryALTinterwordstretchfactor\fontdimen3\font minus
  \fontdimen4\font\relax}
\providecommand{\BIBforeignlanguage}[2]{{%
\expandafter\ifx\csname l@#1\endcsname\relax
\typeout{** WARNING: IEEEtran.bst: No hyphenation pattern has been}%
\typeout{** loaded for the language `#1'. Using the pattern for}%
\typeout{** the default language instead.}%
\else
\language=\csname l@#1\endcsname
\fi
#2}}
\providecommand{\BIBdecl}{\relax}
\BIBdecl

\bibitem{Fu2019Intelligent}
M.~Fu, Y.~Zhou, and Y.~Shi, ``Intelligent reflecting surface for downlink
  non-orthogonal multiple access networks,'' in \emph{Proc. IEEE Global Commun.
  Conf. (Globecom) Workshops}, Waikoloa, Hawaii, USA, Dec. 2019. Available:
  https://arxiv.org/abs/1906.09434.

\bibitem{Andrews2014What}
J.~G. Andrews, S.~Buzzi, W.~Choi, S.~V. Hanly, A.~Lozano, A.~C. Soong, and
  J.~C. Zhang, ``What will 5{G} be?'' \emph{IEEE J. Sel. Areas Commun.},
  vol.~32, no.~6, pp. 1065--1082, Jun. 2014.

\bibitem{Letaief2019roadmap}
K.~B. Letaief, W.~Chen, Y.~Shi, J.~Zhang, and Y.~A. Zhang, ``The roadmap to
  6{G} - {AI} empowered wireless networks,'' \emph{IEEE Commun. Mag.}, vol.~57,
  no.~8, pp. 84--90, Aug. 2019.

\bibitem{Liu2017Nonorthogonal}
Y.~{Liu}, Z.~{Qin}, M.~{Elkashlan}, Z.~{Ding}, A.~{Nallanathan}, and
  L.~{Hanzo}, ``Nonorthogonal multiple access for {5G} and beyond,''
  \emph{Proc. IEEE}, vol. 105, no.~12, pp. 2347--2381, Dec. 2017.

\bibitem{Dai2015Nonorthogonal}
L.~{Dai}, B.~{Wang}, Y.~{Yuan}, S.~{Han}, C.~{I}, and Z.~{Wang},
  ``Non-orthogonal multiple access for 5{G}: {S}olutions, challenges,
  opportunities, and future research trends,'' \emph{IEEE Commun. Mag.},
  vol.~53, no.~9, pp. 74--81, Sept. 2015.

\bibitem{Zhou2018Coverage}
Y.~Zhou, V.~W. Wong, and R.~Schober, ``Coverage and rate analysis of millimeter
  wave {NOMA} networks with beam misalignment,'' \emph{IEEE Trans. Wireless
  Commun.}, vol.~17, no.~12, pp. 8211--8227, Dec. 2018.

\bibitem{Islam2017Power}
S.~M.~R. {Islam}, N.~{Avazov}, O.~A. {Dobre}, and K.~{Kwak}, ``Power-domain
  non-orthogonal multiple access ({NOMA}) in 5{G} systems: Potentials and
  challenges,'' \emph{IEEE Commun. Surveys Tuts.}, vol.~19, no.~2, pp.
  721--742, SecondQuarter 2017.

\bibitem{Ding2017application}
Z.~Ding, Y.~Liu, J.~Choi, Q.~Sun, M.~Elkashlan, and H.~V. Poor, ``Application
  of non-orthogonal multiple access in {LTE and 5G} networks,'' \emph{IEEE
  Commun. Mag.}, vol.~55, no.~2, pp. 185--191, Feb. 2017.

\bibitem{Ding2016MIMONOMA}
Z.~{Ding}, L.~{Dai}, and H.~V. {Poor}, ``{MIMO}-{NOMA} design for small packet
  transmission in the internet of things,'' \emph{IEEE Access}, vol.~4, pp.
  1393--1405, Apr. 2016.

\bibitem{Zhou2018Dynamic}
Y.~{Zhou}, V.~W.~S. {Wong}, and R.~{Schober}, ``Dynamic decode-and-forward
  based cooperative {NOMA} with spatially random users,'' \emph{IEEE Trans.
  Wireless Commun.}, vol.~17, no.~5, pp. 3340--3356, May 2018.

\bibitem{Ding2016UserPairing}
Z.~Ding, P.~Fan, and V.~Poor, ``Impact of user pairing on {5G} non-orthogonal
  multiple access downlink transmissions,'' \emph{IEEE Trans. Veh. Technol.},
  vol.~65, no.~8, pp. 6010--6023, Sept. 2016.

\bibitem{Lu2014Connected}
N.~{Lu}, N.~{Cheng}, N.~{Zhang}, X.~{Shen}, and J.~W. {Mark}, ``Connected
  vehicles: Solutions and challenges,'' \emph{IEEE Internet of Things J.},
  vol.~1, no.~4, pp. 289--299, Aug. 2014.

\bibitem{Cui2014Coding}
T.~J. Cui, M.~Q. Qi, X.~Wan, J.~Zhao, and Q.~Cheng, ``Coding metamaterials,
  digital metamaterials and programmable metamaterials,'' \emph{Light, Sci.
  Appl.}, vol.~3, no.~10, p. 218, 2014.

\bibitem{Di2019Smart}
M.~Di~Renzo, M.~Debbah, D.-T. Phan-Huy \emph{et~al.}, ``Smart radio
  environments empowered by reconfigurable {AI} meta-surfaces: {A}n idea whose
  time has come,'' \emph{EURASIP J. Wireless Commun. Netw.}, May 2019.

\bibitem{Yuan2020Reconfigurable}
X.~Yuan, Y.-J. Zhang, Y.~Shi, W.~Yan, and H.~Liu,
  ``Reconfigurable-intelligent-surface empowered {6G} wireless communications:
  {C}hallenges and opportunities,'' \emph{arXiv preprint arXiv:2001.00364},
  2020.

\bibitem{Huang2019Holographic}
C.~Huang, S.~Hu, G.~C. Alexandropoulos, A.~Zappone, C.~Yuen, R.~Zhang,
  M.~Di~Renzo, and M.~Debbah, ``Holographic {MIMO} surfaces for {6G} wireless
  networks: {O}pportunities, challenges, and trends,'' \emph{arXiv preprint
  arXiv:1911.12296}, 2019.

\bibitem{Liang2019Large}
Y.-C. Liang, R.~Long, Q.~Zhang, J.~Chen, H.~V. Cheng, and H.~Guo, ``Large
  intelligent surface/antennas ({LISA}): Making reflective radios smart,''
  \emph{J. Commun. Inf. Networks}, vol.~4, no.~2, pp. 40--50, Jun. 2019.

\bibitem{Wu2019Towards}
Q.~{Wu} and R.~{Zhang}, ``Towards smart and reconfigurable environment:
  Intelligent reflecting surface aided wireless network,'' \emph{IEEE Commun.
  Mag.}, vol.~58, no.~1, pp. 106--112, Jan. 2020.

\bibitem{Shi2014Group}
Y.~{Shi}, J.~{Zhang}, and K.~B. {Letaief}, ``Group sparse beamforming for green
  {C}loud-{RAN},'' \emph{IEEE Trans. Wireless Commun.}, vol.~13, no.~5, pp.
  2809--2823, May 2014.

\bibitem{Huang2019Reconfigurable}
C.~{Huang}, A.~{Zappone}, G.~C. {Alexandropoulos}, M.~{Debbah}, and C.~{Yuen},
  ``Reconfigurable intelligent surfaces for energy efficiency in wireless
  communication,'' \emph{IEEE Trans. Wireless Commun.}, vol.~18, no.~8, pp.
  4157--4170, Aug. 2019.

\bibitem{Wu2019intelligentJ}
Q.~Wu and R.~Zhang, ``Intelligent reflecting surface enhanced wireless network
  via joint active and passive beamforming,'' \emph{IEEE Trans. Wireless
  Commun.}, vol.~18, no.~11, p. 5394–5409, Aug. 2019.

\bibitem{Guo2019Weighted}
H.~{Guo}, Y.~{Liang}, J.~{Chen}, and E.~G. {Larsson}, ``Weighted sum-rate
  maximization for reconfigurable intelligent surface aided wireless
  networks,'' \emph{IEEE Trans. Wireless Commun.}, Feb. 2020, doi
  10.1109/TWC.2020.2970061.

\bibitem{Jiang2019Over}
T.~Jiang and Y.~Shi, ``Over-the-{A}ir computation via intelligent reflecting
  surfaces,'' in \emph{Proc. IEEE Global Commun. Conf. (Globecom)}, Waikoloa,
  Hawaii, USA, Dec. 2019.

\bibitem{Yu2019Enabling}
X.~Yu, D.~Xu, and R.~Schober, ``Enabling secure wireless communications via
  intelligent reflecting surfaces,'' in \emph{Proc. IEEE Global Commun. Conf.
  (Globecom)}, Waikoloa, Hawaii, USA, Dec. 2019.

\bibitem{Chen2019Intelligent}
J.~Chen, Y.-C. Liang, Y.~Pei, and H.~Guo, ``Intelligent reflecting surface: A
  programmable wireless environment for physical layer security,'' \emph{arXiv
  preprint arXiv:1905.03689}, 2019.

\bibitem{Xia2019Intelligent}
S.~Xia and Y.~Shi, ``Intelligent reflecting surface for massive device
  connectivity: Joint activity detection and channel estimation,'' in
  \emph{Proc. of EEE Int. Conf. Acoust. Speech Signal Process. (ICASSP)},
  Barcelona, Spain, May 2020.

\bibitem{Ding2019simple}
Z.~Ding and H.~V. Poor, ``{A simple design of {IRS}-{NOMA} transmission},''
  \emph{arXiv preprint arXiv:1907.09918}, 2019.

\bibitem{Yang2019Intelligent}
G.~Yang, X.~Xu, and Y.-C. Liang, ``Intelligent reflecting surface assisted
  non-orthogonal multiple access,'' \emph{arXiv preprint arXiv:1907.03133},
  2019.

\bibitem{Li2019Joint}
Y.~Li, M.~Jiang, Q.~Zhang, and J.~Qin, ``Joint beamforming design in
  multi-cluster {MISO} {NOMA} intelligent reflecting surface-aided downlink
  communication networks,'' \emph{arXiv preprint arXiv:1909.06972}, 2019.

\bibitem{Zhu2019Power}
J.~Zhu, Y.~Huang, J.~Wang, K.~Navaie, and Z.~Ding, ``Power efficient
  {IRS}-assisted {NOMA},'' \emph{arXiv preprint arXiv:1912.11768}, 2019.

\bibitem{Mu2019exploiting}
X.~Mu, Y.~Liu, L.~Guo, J.~Lin, and N.~Al-Dhahir, ``Exploiting intelligent
  reflecting surfaces in multi-antenna aided {NOMA} systems,'' \emph{arXiv
  preprint arXiv:1910.13636}, 2019.

\bibitem{Zuo2020Resource}
J.~Zuo, Y.~Liu, S.~Member, and Z.~Qin, ``Resource allocation in intelligent
  reflecting surface assisted {NOMA} systems,'' \emph{arXiv preprint
  arXiv:2002.01765}, 2020.

\bibitem{Nadeem2019Intelligent}
Q.-U.-A. Nadeem, A.~Kammoun, A.~Chaaban, M.~Debbah, and M.-S. Alouini,
  ``Intelligent reflecting surface assisted multi-user {MISO} communication,''
  \emph{arXiv preprint arXiv:1906.02360}, 2019.

\bibitem{Yang2019IntelligentOFDM}
Y.~Yang, B.~Zheng, S.~Zhang, and R.~Zhang, ``Intelligent reflecting surface
  meets {OFDM}: Protocol design and rate maximization,'' \emph{arXiv preprint
  arXiv:1906.09956}, 2019.

\bibitem{He2019Cascaded}
Z.~{He} and X.~{Yuan}, ``Cascaded channel estimation for large intelligent
  metasurface assisted massive {MIMO},'' \emph{IEEE Wireless Commun. Lett.},
  2019.

\bibitem{Taha2019Enabling}
A.~Taha, M.~Alrabeiah, and A.~Alkhateeb, ``Enabling large intelligent surfaces
  with compressive sensing and deep learning,'' in \emph{Proc. IEEE Global
  Commun. Conf. (Globecom)}, Waikoloa, Hawaii, USA, Dec. 2019.

\bibitem{Liu2018Multiple}
Y.~{Liu}, H.~{Xing}, C.~{Pan}, A.~{Nallanathan}, M.~{Elkashlan}, and
  L.~{Hanzo}, ``Multiple-antenna-assisted non-orthogonal multiple access,''
  \emph{IEEE Wireless Commun.}, no.~2, pp. 17--23, Apr. 2018.

\bibitem{Alavi2018Beamforming}
F.~Alavi, K.~Cumanan, Z.~Ding, and A.~G. Burr, ``Beamforming techniques for
  nonorthogonal multiple access in {5G} cellular networks,'' \emph{IEEE Trans.
  Veh. Technol.}, vol.~67, no.~10, pp. 9474--9487, Oct. 2018.

\bibitem{Al2019Energy}
H.~Al-Obiedollah, K.~Cumanan, J.~Thiyagalingam, A.~G. Burr, Z.~Ding, and O.~A.
  Dobre, ``Energy efficient beamforming design for {MISO} non-orthogonal
  multiple access systems,'' \emph{IEEE Trans. Commun.}, vol.~67, no.~6, pp.
  4117--4131, Jun. 2019.

\bibitem{Zhu2020optimal}
J.~{Zhu}, J.~{Wang}, Y.~{Huang}, K.~{Navaie}, Z.~{Ding}, and L.~{Yang}, ``On
  optimal beamforming design for downlink {MISO} {NOMA} systems,'' \emph{IEEE
  Trans. Veh. Technol.}, Jan. 2020, doi 10.1109/TVT.2020.2966629.

\bibitem{Bjornson2014optimal}
E.~{Bj{\"o}rnson}, M.~{Bengtsson}, and B.~{Ottersten}, ``Optimal multiuser
  transmit beamforming: {A} difficult problem with a simple solution
  structure,'' \emph{IEEE Signal Process. Mag.}, vol.~31, no.~4, pp. 142--148,
  Jul. 2014.

\bibitem{Ma2010Semidefinite}
Z.~{Luo}, W.~{Ma}, A.~M. {So}, Y.~{Ye}, and S.~{Zhang}, ``Semidefinite
  relaxation of quadratic optimization problems,'' \emph{IEEE Signal Process.
  Mag.}, vol.~27, no.~3, pp. 20--34, May 2010.

\bibitem{Chen2017ADMM}
E.~Chen and M.~Tao, ``{ADMM}-based fast algorithm for multi-group multicast
  beamforming in large-scale wireless systems,'' \emph{IEEE Trans. Commun.},
  vol.~65, no.~6, pp. 2685--2698, Jun. 2017.

\bibitem{Yang2019Data}
K.~Yang, Y.~Shi, and Z.~Ding, ``Data shuffling in wireless distributed
  computing via low-rank optimization,'' \emph{IEEE Trans. Signal Process.},
  vol.~67, no.~12, pp. 3087--3099, Jun. 2019.

\bibitem{Yang2020Federated}
K.~{Yang}, T.~{Jiang}, Y.~{Shi}, and Z.~{Ding}, ``Federated learning via
  over-the-air computation,'' \emph{IEEE Trans. Wireless Commun.}, Jan. 2020,
  doi 0.1109/TWC.2019.2961673.

\bibitem{Dinh1997d.c}
P.~D. Tao and L.~T.~H. An, ``Convex analysis approach to {DC} programming:
  Theory, algorithms and applications,'' \emph{Acta Math. Vietnam.}, vol.~22,
  no.~1, pp. 289--355, 1997.

\bibitem{Rockafellar2015Convex}
R.~T. Rockafellar, \emph{Convex Analysis}.\hskip 1em plus 0.5em minus
  0.4em\relax Princeton university press, 2015.

\bibitem{Dong2018demixing}
J.~{Dong} and Y.~{Shi}, ``Nonconvex demixing from bilinear measurements,''
  \emph{IEEE Trans. Signal Process.}, vol.~66, no.~19, pp. 5152--5166, Oct.
  2018.

\bibitem{grant2014cvx}
M.~Grant and S.~Boyd, ``{CVX}: Matlab software for disciplined convex
  programming, version 2.1,'' Mar. 2014.

\bibitem{Zhou2018Stable}
Y.~Zhou, V.~W. Wong, and R.~Schober, ``Stable throughput regions of
  opportunistic {NOMA} and cooperative {NOMA} with full-duplex relaying,''
  \emph{IEEE Trans. Wireless Commun.}, vol.~17, no.~8, pp. 5059--5075, May
  2018.

\bibitem{Wu2019BeamformingDiscrete}
Q.~{Wu} and R.~{Zhang}, ``Beamforming optimization for wireless network aided
  by intelligent reflecting surface with discrete phase shifts,'' \emph{IEEE
  Trans. Commun.}, vol.~68, no.~3, pp. 1838--1851, Mar. 2020.

\bibitem{Griffin2009Complete}
J.~D. {Griffin} and G.~D. {Durgin}, ``Complete link budgets for
  backscatter-radio and {RFID} systems,'' \emph{IEEE Antennas Propag. Mag.},
  vol.~51, no.~2, pp. 11--25, Apr. 2009.

\end{thebibliography}

\end{document}